%% file: sud-v2.tex
\documentclass[12pt]{article}
\usepackage{wor-sci,graphicx}

\input{mclmath}
\renewcommand{\thefootnote}{\fnsymbol{footnote}}
\def\eqref#1{#1}
\def\er(#1){(\eqref{#1})}
\def\eq(#1){eq.~\er(#1)}
\def\eqalno(#1){(#1)}

\def\defref{\item\label}
\def\rf#1#2#3{{\bf #1}, #2 (19#3)}

\def\deffigA#1#2#3{\begin{figure} 
    \centering
    \includegraphics{#2}
    \caption{#3}
    \end{figure}
}
\def\deffigB#1#2#3{\begin{figure} 
    #2
    \caption{#3}
    \end{figure}
}
\def\fig#1{Fig.~#1}
\def\Lag{{\cal L}}
\def \jmu {\bar q \gamma^\mu q}
\def \gcusp {\Gamma_{\rm cusp}}
\def \msoft {{\cal M}_{\rm soft}}

\worcite
\begin{document} 
% \rightline{\vbox{\halign{&#\hfil\cr
%  &\normalsize ANL-HEP-PR-84-36\cr
%  \normalsize 2 Feb 89.  Printed \today\cr}}}
%  \vspace{1in}
\begin{center}

% \Large
{\bf SUDAKOV FORM FACTORS}\footnote{%
The original version of this paper was published in: ``Sudakov form
factors", in ``Perturbative QCD" (A.H. Mueller, ed.) (World
Scientific, Singapore, 1989).  This version differs by an update of the
author's address to the current one, and the correction of some misprints
and other minor errors.}
\medskip

\normalsize John C. Collins
\\ \smallskip
        Physics Department,\\
        Penn State University,\\
        University Park PA 16802, \\
        U.S.A. 
\\\medskip
\end{center}

\medskip

\begin{abstract}
  The theory of the on-shell Sudakov form factor to all orders of logarithms is
explained.
\end{abstract}

\renewcommand{\thefootnote}{\arabic{footnote}} 

\sec {Introduction}

The key to understanding and using perturbative QCD is the idea of
factorization. Factorization is the property that some cross-section or
amplitude is a product of two (or more) factors and that each factor
depends only on physics happening on one momentum (or distance) scale.
The process is supposed to involve some large momentum transfer, on a
scale $Q$, and corrections to the factorized form are suppressed by a
power of $Q$. (In general the product is in the sense of a
matrix product or of a convolution.)

The standard factorization theorems are typified by the one for the
moments of the deep inelastic structure functions:
$$
F_n(Q) = C_n \left(\al(Q)\right) \ast \exp\left[\int^Q_{Q_0}\gamma_n
\left(\alpha_s(\mu)\right) \d\mu/\mu\right] \ast M_n(Q_0).\eqno(1)
$$
Here $F_n$ is the $n$th moment of one of the structure functions,
$C_n$ is a Wilson coefficient, $M_n$ is a hadronic matrix element of
an operator of spin $n$ and twist 2, 
$\gamma_n$ is an anomalous dimension, and $Q_0$ is a fixed
scale.  The symbol `$\ast$' denotes a matrix product (to allow the
possibility of contributions from more than one operator).  
The renormalization
group has been used to absorb all logarithms of large mass ratios into the
integral over the anomalous dimension.  These theorems are described
elsewhere in this volume.  

In this article, I will treat the Sudakov
form factor, which provides the simplest example of factorization
theorems of a more complicated kind.  The difference between this
case and deep inelastic scattering results from a difference
in the regions of loop-momentum space that give the
leading-twist contributions to the process. In the case of simple
factorization theorems like \er(1), these regions involve lines with
momenta that are either collinear to the detected particles or are
far off-shell.  In individual graphs there are leading twist contributions
from regions with soft gluons; but after an intricate cancellation\cite{1,2},
the effects of soft gluons cancel.  However, in the Sudakov form
factor the effects of the soft gluons do not cancel.  Even so, 
a more general factorization theorem holds for this case.  

This form factor is the elastic form factor of an elementary
particle in an abelian 
gauge theory at large momentum transfer $Q$.  Sudakov\cite{3}
treated the off-shell form factor in the leading logarithmic
approximation.  I will treat the on-shell case and derive the full
factorization formula, which is valid to all orders of perturbation
theory and includes all nonleading logarithms.

This relatively simple case is a prototype for such processes as the
Drell-Yan cross section when the transverse momentum is much less than
the invariant mass of the Drell-Yan pair.  A treatment to all orders of
logarithms is given in \cite{4,5}, and applications to phenomenology can be
found in \cite{6,7}.  Work at the leading logarithm level can be found in
\cite{8,9} and references therein.  In \cite{10} Sen showed how to treat
the on-shell Sudakov form factor in a non-abelian theory to all orders of
logarithms.

The first step in proving any factorization theorem is to understand the
regions of the space of loop momenta that give the ``leading-twist''
contributions, i.e. contributions not suppressed by a power of $Q$.
After appropriate approximations it is possible to
use Ward identities to convert the leading-twist contributions into a
form that corresponds to the factorization theorem.  A complication here is
to eliminate double counting.  Finally a differential equation
for the evolution of the form factor is derived. It is only after this
step that it is possible to perform systematic perturbative
calculations, without having the validity of a finite-order calculation
being brought into doubt by the possibility of large logarithmic
corrections in higher order.  The solution to the equation is in
terms of quantities with perturbation expansions that have no
large logarithmic terms in their coefficients.

One topic I will emphasize is the extent to which the fact that we are
dealing with a renormalizable gauge theory of physics comes into the
form of the factorization. To do this I will start by examining what
sort of result holds in a superrenormalizable theory without gauge
fields, specifically $\phi^3$ theory in four space-time dimensions. The
$\left(\phi^3\right)_4$ theory is of course completely unphysical.
However, it is a simple model which exhibits the features common to any
superrenormalizable theory without gauge fields but which has no
irrelevant complications. We will see that a very simple factorization
holds true.

Next, I will step up the space-time dimension to $d=6$. This will render
the model merely renormalizable instead of superrenormalizable. The
short-distance part of the factorization will then become non-trivial,
but it will still be of the same form as for deep inelastic scattering,
\er(1).

Finally, I will return to four dimensions, but now with a gauge theory.
To provide a simple demonstration of how and why all the logarithms are
under control, I must avoid complications that are irrelevant for this
purpose. So I will take the theory to be abelian, with a massive gluon,
and treat the annihilation of a $q\bar q$ pair into a virtual photon.
The complications thereby avoided include: a non-abelian
gauge group, the infra-red divergences caused by a massless gluon (which
would mean we would have to discuss a kinematically more complicated
process), and color confinement (which would force us to treat, say, a
form factor of a composite particle). Although these complications are
important for real strong interactions, they are inessential if we are
trying to understand ``Sudakov'' effects by themselves.

\sec {Reduced graphs}

Consider a form factor                 
$$
F=\langle 0|j(0)|p_Ap_B\rangle.  \eqno(2)
$$
Here $j$ is a composite field, for example the electromagnetic current
of quarks in QCD, and $|p_Ap_B\rangle$
represents an incoming two-particle state with energy $Q$:
$$
Q^2=\left(p_A+p_B\right)^2,\eqno(3)
$$
which we assume to be very large. 

First we must find the
regions of momentum space that are important in Feynman graphs for this
amplitude. We use the method given by Libby
and Sterman\cite{11}.  Suppose we scale all momenta by a factor $Q$:
$$
\eqalign{
k^\mu&=\tilde k^\mu Q,\cr
m&=\tilde m Q.}\eqno(4)
$$
The reduced mass $\tilde m$ goes to zero as $Q$ goes to infinity, so
that we are effectively going to a massless theory.  If all scaled
momenta in a graph are off-shell by order unity, then we get a
contribution of order unity (given that we are in a renormalizable theory,
so that the coupling is dimensionless).  
Then simple perturbation theory is
applicable provided only that the effective coupling $\al(Q)$ is small.
Other leading 
contributions can come from regions where some of the scaled momenta
become on-shell in the massless theory $(\tilde m=0)$. In this case we
obtain a contribution only when the contour of integration is trapped at
the on-shell point, for otherwise we may deform the contours into the
off-shell region.  Such points were called pinch-singular points
in \cite{11}.

The analysis of Coleman and Norton\cite{12} can be used to locate the
pinch singular points. Each pinch-singular point can be represented by a
reduced graph. The lines whose scaled momenta are off-shell by order
unity are all contracted to points; they form the vertices of the
reduced graph for a pinch singular point. The lines that have on-shell
scaled momenta form the lines of the reduced graph. In order that the
contours of integration be pinched, the reduced graph must represent a
classical scattering process.  In the case of an annihilation form
factor, the reduced graphs have the form exemplified by \fig{1}. The
on-shell lines either have non-zero fractions of the scaled momentum of
one or other of the incoming lines or they have zero scaled momentum.
These lines are represented by solid and dashed lines (respectively),
and are called jet and soft lines.  An arbitrary number of jet lines
parallel to $\tilde p^\mu_A$ interact and enter the reduced vertex where
the annihilation occurs. A similar situation occurs for $\tilde p_B$. An
arbitrary number of soft lines join the two jet subgraphs.

\deffigA{1}{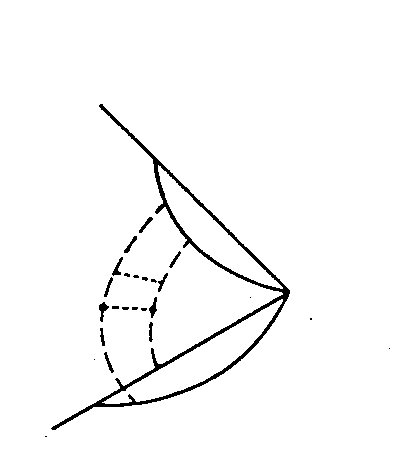}{Typical reduced graph for annihilation form factor.  Jet
lines are solid, and soft lines are dashed.  }

This can all be said without knowing the field theory. We next need
to know which of the pinch singular points give important contributions
as $Q \to \infty$.  For this purpose we consider only leading twist
contributions, i.e., those that are not suppressed by a
power of $Q$.  Which regions give
leading-twist contributions will depend on the theory within which we
work, especially on its renormalizability or superrenormalizability and
on the presence or absence of gauge particles.

\sec {Superrenormalizable Scalar Theory: $\left(\phi^3\right)_4$}

The Lagrangian of $\phi^3$ theory is
$$
\Lag
={1\over 2}\left(\partial\phi\right)^2-{1\over 2}m^2_B\phi^2-{1\over 6}
g\phi^3. \eqno(5)
$$
In $d=4$ space-time dimensions the coupling, $g$, has positive mass
dimension. This signals that no infinite coupling or wave function
renormalization is needed, i.e., that the theory is superrenormalizable.
We define the form factor in \eq(2) by choosing the composite field $j$
to be $\half \phi^2$.

\deffigB{2}{%
   \centering
   \def\tmpscale{0.95}
   \includegraphics[scale=\tmpscale]{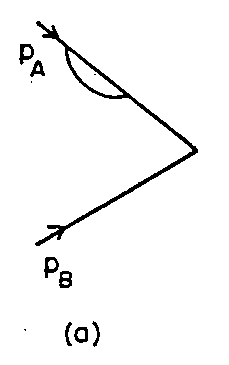}
   \includegraphics[scale=\tmpscale]{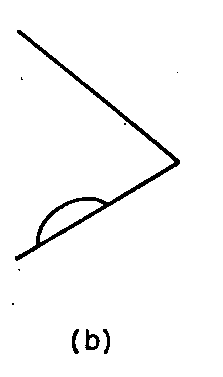}
   \includegraphics[scale=\tmpscale]{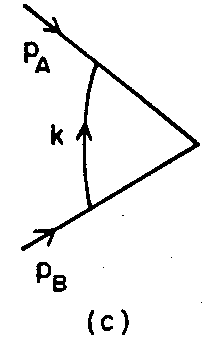}
}{One loop graphs for form factor in $\phi^3$ theory.}

First consider the one-loop graphs, which are listed in \fig{2}. The
graphs with
self-energy corrections, \fig{2}(a) and (b), have reduced graphs equal
to themselves. If these and higher-order self-energy graphs were all that
we have, then the form factor would be equal to
$$
z(m^2,g), \eqno(6)
$$
where $z$ is the residue of the renormalized propagator:
$$
S_F(p^2,m^2,g) \to {i z\over p^2-m^2} \quad \hbox{as $p^2 \to m^2$}.
\eqno(7)
$$

In fact these graphs are all that we have, for vertex graphs like \fig{2}(c)
vanish as $Q\to \infty$, by a power $Q$, as we will now show.
\fig{2}(c) has the value
$$ 
\eqalignno{
   \Gamma_c&={-ig^2\over (2\pi)^4}\int \d{^4k} {1\over (m^2-k^2)
   \left[m^2 - (p_A+k)^2 \right] \left[m^2 - (p_B-k)^2 \right]}
   &\eqalno(8)
\cr                                   
   &={-ig^2\over (2\pi)^4Q^2} \int \d{^4\tilde k} {1\over
    (\tilde m^2  - \tilde k^2)
    \left[ \tilde m^2 - (\tilde p_A + \tilde k)^2 \right]
    \left[\tilde m^2 - (\tilde p_B - \tilde k)^2 \right]}
\cr
   &&\eqalno(9)
}
$$
The possible reduced graphs for \fig{2}(c) are listed in \fig{3}:
\begin{enumerate}
\item[(a)] \fig{3}(a) corresponds to the region that $|\tilde k^\mu|\sim 1$,
i.e.~where all internal lines are far off-shell. Manifestly the resulting
contribution is $O\left(1/Q^2\right)$. This is essentially a result of
dimensional analysis coupled with the positive dimension of $g$.
\item[(b)] \fig{3}(b) corresponds to the region where $k^\mu$ is
collinear to $p_A^\mu$.  This graph is also $O\left(1/Q^2\right)$. To
see this, we use light cone coordinates
$$
p^-_B=p^+_A={Q\over \sqrt 2},\qquad p^+_B=p^-_A={m^2\over Q\sqrt 2}.
\eqno(10)
$$
Then in the region symbolized by \fig{3}(b), we have
$$
k^+=O(Q),\quad k^-=O(\lambda^2Q),\quad k_T=O(\lambda Q),\eqno(11)
$$
where $\lambda$ is small.  It is now easy to check that the contribution to
the form factor is $O\left(1/Q^2\right)$.  The point is that one quark line
is far off-shell and that there are no compensating numerator factors.
\item[(c)] \fig{3}(c) is just \fig{3}(b) with $A\leftrightarrow B$.
\item[(d)] For \fig{3}(d), which corresponds to the region where all
components of $k^\mu$ are much less than $Q$, we let
$$
k^\mu=O\left(\lambda Q\right),
$$
for all components. Again we get a contribution of order $1/Q^2$.
\end{enumerate}

\deffigB{3}{%
   \centering
   \def\tmpscale{0.85}
   \includegraphics[scale=\tmpscale]{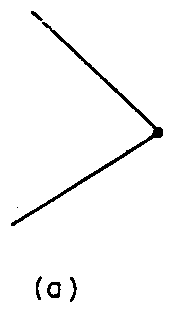}
   \includegraphics[scale=\tmpscale]{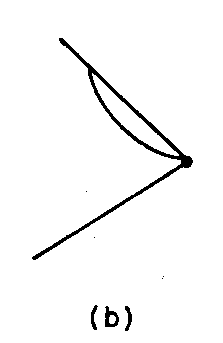}
   \includegraphics[scale=\tmpscale]{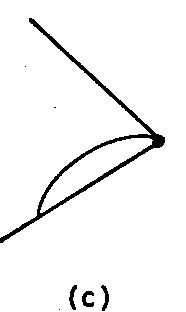}
   \\
   %\vphantom{\includegraphics[scale=\tmpscale]{Fig3c}}
   \includegraphics[scale=\tmpscale]{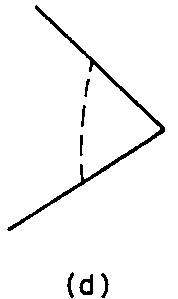}
}{Reduced graphs for \fig{2}(c).}

It is fairly easy to show that this analysis holds true to all
orders, by the methods of \cite{11}.  The reduced graphs for
the leading twist contributions all have the form of self energy
graphs attached to the lowest order vertex, so that the form factor
is $z$ plus higher-twist contributions.  (See \eq(7) for
the definition of $z$.)

\sec {Renormalizable scalar theory: $(\phi^3)_6$}

\subsec {One-loop}

The sole significant difference in going to six space-time dimensions is
caused by the coupling's becoming dimensionless and the consequent
need for coupling and wave function renormalization.  We write the
Lagrangian in the form:
$$
\eqalign{
  \Lag ={}&{1\over 2}(\partial\phi)^2-{1\over 2}m^2\phi^2-{1\over
  6} \mu^{\epsilon } g\phi^3
\cr
  &+{1\over 2}\delta Z(\partial\phi)^2-{1\over 2}\delta m^2\phi^2-{1\over 6}
  \mu^{\epsilon } \delta g\phi^3 + h \phi.
}
\eqno(12)
$$
Here $m$ and $g$ are the renormalized mass and coupling, and the last four
terms are the renormalization counterterms. We will use dimensional
regularization (i.e., space-time dimension
$d=6-2 \epsilon$) to cut off the ultra-violet
divergences.  To keep the coupling $g$ dimensionless we
introduce the unit of mass\cite{13} $\mu$. The linear 
term $h\phi$ is
adjusted to cancel tadpole graphs; for the other terms we will use
\MSbar\ renormalization\cite{14}.

The structure of the reduced graphs is the same, as always. What changes
is the size of the contributions. Consider the one-loop vertex graph
\fig{2}(c), whose reduced graphs are in \fig{3}. By following the same
method as we used in Sec.~3, we find that the contributions of the reduced
graphs (at $d=6$) are
$$
\eqalign{
\hbox{\fig{3}(a)}\qquad&Q^0,\cr
\hbox{\fig{3}(b) or (c)}\qquad&\lambda^2Q^0,\cr
\hbox{\fig{3}(d)}\qquad&\lambda^2Q^0.}\eqno(13)
$$
Clearly we get a leading contributions solely from the region where all
internal lines are far off-shell. The existence of this contribution is
tied to the dimensionlessness of the coupling.

The contribution of \fig{2}(c) is therefore given by neglecting all masses,
with errors of order $1/Q^2$. Thus
$$
\Gamma_c\sim -{ig^2\over (2\pi)^6}(2\pi\mu)^{2\epsilon }
\int \d{^{6-2\epsilon}k}
{1\over (-k^2)\left[-(\hat p_A+k)^2\right]\left[-(\hat p_B-k)^2\right]}
+{}\hbox{counterterm}.\eqno(14)
$$
Here $\hat p_A^\mu$ and $\hat p^\mu_B$ are light-like vectors close to 
$p_A^\mu$ and $p_B^\mu$:
$$
\eqalign{
  (\hat p^+_A, \hat p^-_A) &= (p^+_A, 0) = (Q/\sqrt 2, 0),\cr
  (\hat p^+_B, \hat p^-_B) &= (0, p^-_B) = (0, Q/\sqrt 2)}
\eqno(15)
$$

After using \MSbar\ renormalization to cancel the ultraviolet divergence
in the integral in \er(14), we find
$$
\Gamma_c={g^2\over 128\pi^3} \left[-\ln(-Q^2/ \mu^2)
+3\right] + O(1/Q^2).
\eqno(16)
$$
(At one-loop order, the \MSbar\ scheme is defined by requiring counterterms
to be a coefficient times $ 1 / \epsilon - \gamma + \ln (4\pi )$, where
$\gamma $ is Euler's constant.  In the MS scheme we would omit the $\gamma$
and the $\ln (4\pi )$.)

\subsec {Higher orders}

\deffigB{4}{%
   \centering
   \includegraphics[scale=0.94]{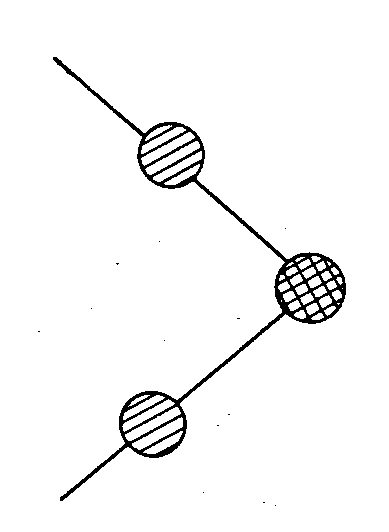}
}{General graph for form factor.  The cross-hatched 
bubble is the sum of all 1PI graphs for the form factor.}

The generalization to all orders of the one-loop results is obtained by
observing that the graphs for the form factor are a product of two propagators
and a one-particle-irreducible (1PI) vertex (\fig{4}). A leading contribution
is only obtained from the 1PI vertex when all its internal lines are off-shell
by order $Q^2$.  It can be shown fairly easily that other regions
are power suppressed\cite{15}.  
Thus the only  leading reduced graphs have the form of
\fig{5}.  This result is true in any renormalizable non-gauge theory.

\deffigB{5}{%
   \centering
   \includegraphics[scale=0.95]{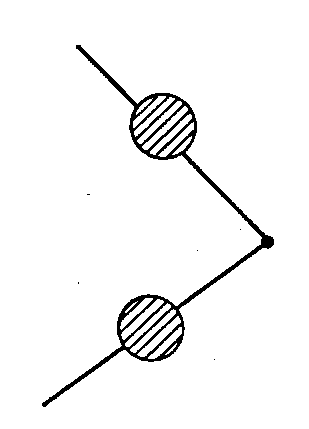}
}{Leading reduced graphs for form factor in $(\phi^3)_6$
theory.  }

Therefore the form factor has the form
$$
F = z(g,m/\mu) \, \Gamma(Q/\mu,g) +
O\left((\hbox{logarithms of $Q$}) /Q^2\right).\eqno(17)
$$
Here $\Gamma$ is the 1PI vertex with the masses set to zero.  
This is the simplest example of a factorization theorem. The $z$ factor
comes from the single-particle propagator; it depends on phenomena on the
scale of the quark 
mass $m$ and is independent of the energy $Q$. On the other
hand the vertex factor $\Gamma$ depends only on the total energy and not
on the mass.

Since the theory needs renormalization, there is important dependence on
the unit of mass $\mu$. A perturbative calculation of $z$ or $\Gamma$
has large logarithms of $m/\mu$ or of $Q/\mu$, so a simultaneous 
direct calculation
of these quantities to low order 
cannot be reliable if $Q/m$ is large enough. The
renormalization group comes to our aid since both $z$ and $\Gamma$
satisfy renormalization group equations\cite{16,17}
$$
\eqalignno{
   \mu{\d{}\over \d\mu} z&=-2\gamma\left(g\left(\mu\right)\right)
&\eqalno(18)\cr
   \mu{ \d{} \over \d\mu} \Gamma&=-\gamma_{\phi^2}
        \left(g\left(\mu\right)\right)+2\gamma
        \left(g\left(\mu\right)\right)
\cr
   &\equiv -\hat\gamma\left(g\left(\mu\right)\right).
&\eqalno(19)\cr
}
$$
Here $\gamma(g)$ and $\gamma_{\phi^2}(g)$ are the anomalous dimensions of
the operators $\phi$ and $\phi^2$ respectively, and the renormalization-group
operator is
$$
\mu{\d{}\over \d\mu} \equiv
\mu{\partial\over \partial\mu}+\beta(g){\partial\over \partial
g}+\gamma_{\phi^2}\, m^2{\partial\over \partial m^2}.\eqno(20)
$$
Note that $\gamma$ is half its value as defined by many authors.

Evidently we can solve Eqs.~\er(18) and \er(19) and write
% For v.2: Brackets corrected.
$$
\eqalign{
    F(Q,m,g,\mu)={} &z(g(m),1) \,
      \exp \bigg[2\int^m_\mu {\d{\mu'}\over \mu'} \gamma\left(g(\mu')\right)
           \bigg] \cr
   &\times \exp \bigg[ \int^Q_\mu {\d{\mu'}\over \mu'}
          \hat\gamma \left(g(\mu' )\right)
           \biggr]\,
          \Gamma (1, g(Q)).}
\eqno(21)
$$
Here $g(\mu)$ is the running (or effective) coupling at scale $\mu$. Evidently
each factor may be reliably calculated without large logarithms in higher-order
corrections.

\subsec {``Optimization'' of perturbation calculations}

In \er(21) the endpoints of the integrals
over $\mu'$ are $\mu'=m$ and $\mu'=Q$.  This is not necessary;
all that is needed is that the endpoints be of {\it order} $m$ and $Q$.
This is important in
``optimizing'' perturbative calculations. We can write
$$
\eqalign{
  F(Q,m,g,\mu) = z\left(g(c_1m),1/c_1\right)
  & \exp
    \left[2\int_\mu^{c_1m} {\d{\mu'}\over \mu'} \, \gamma\left(g(\mu')\right)
    \right]
\cr
   & \times \exp \left[\int_\mu^{c_2Q} {\d{\mu'}\over \mu'} \,
       \hat\gamma\left(g(\mu') \right)
    \right]
    \, \Gamma\left(1/c_2, g(c_2Q) \right).
}
\eqno(22)
$$
Here $c_1$ and $c_2$ are arbitrary constants to be chosen at will. In a
calculation to all orders of perturbation theory the result for $F$ is
independent of our choice of $c_1$ and $c_2$.  But in a finite order
calculation the result has dependence on $c_1$ and $c_2$ of the order of
the first uncalculated term.  We should choose $c_1$ and $c_2$ not too
far from unity to keep higher order corrections small. The change in
$\Gamma$ given by varying $c_1$ and $c_2$ by a factor of 2 can be
regarded as an estimate of the error in the calculation induced by
uncalculated higher order terms.

There has been much discussion\cite{18} of appropriate ways to choose $c_1$
and $c_2$.

\sec {Gauge theories}
\setcounter{footnote}{0}

We now consider a form factor in the massive abelian gauge theory whose
Lagrangian is
$$
\eqalign{
  \Lag = {}&-{1\over 4}F^2_{\mu\nu} + {1\over 2}m^2A^2 -
     {1\over 2\xi}\partial\cdot A^2
   +\bar q\left(i\st\partial+\mu^{\epsilon }g\st A-M\right)q
\cr
   &+\hbox{UV counterterms}.
}\eqno(23)
$$
Here the notation is standard. The renormalized masses and coupling are
$m$, $M$ and $g$. We regulate ultra-violet divergences by continuing to
space-time dimension $d=4-2\epsilon$, and we will call $A_\mu$ the gluon field
and $q$ the quark field.
We will treat the electromagnetic form factor of the 
quark\footnote{To be precise, note
that the operator $[\jmu]$ in \er(24) is the renormalized operator
$\jmu + \hbox{UV counterterms}$.}
$$
F = \left< 0\left|\, [\jmu](0) \, \right|q(p_A),\bar q(p_B)\right>,
\eqno(24)
$$
when the center-of-mass energy 
$Q$ \hbox{$\left(\equiv \sqrt{(p_A+p_B)^2}\right)$}
gets large compared to all masses.  

Sudakov\cite{3} was the first to discuss such a form factor to all orders
of perturbation theory. His result was for the sum of all the leading
logarithms, but with the quarks off-shell
% For v.2 
and a massless photon.  
He found that
$F \sim \exp \left[-(g^2 / 8\pi^2) \ln^2(Q^2)\right]$.
The
on-shell case, with a massive photon, was first treated
(still in leading logarithm approximation) by Jackiw\cite{19}, with
the result that $F \sim \exp \left[-(g^2 / 16\pi^2) \ln^2(Q^2)\right]$.

From Sudakov's work (1956) until 1980, there was no progress in going
systematically beyond leading logarithms, despite many attempts.
Mueller\cite{20} and Collins\cite{21} then gave an all-orders and
all-logarithms treatment. The treatment below is an improved version 
of \cite{21}.  A first version of the present treatment
appeared\cite{22} as notes on lectures given in 1984.

Notice that in these gauge theory form factors 
there are two logarithms of $Q$ per loop
rather than the one logarithm per loop that we have in $(\phi^3)_6$ theory.
This is a symptom of the new physics present in a gauge theory. The effects
that we will investigate reappear in many processes in QCD.

\subsec {One loop}

\deffigB{6}{%
   \centering
   \includegraphics[scale=0.95]{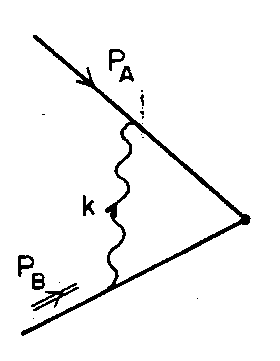}
}{One loop vertex in QED.  }

Self-energy graphs contribute just as they do in $\phi^3$ theory, and give
an overall factor 
$z_2(g,m,M,\mu)$, the residue of the pole of the quark propagator. 
So the only non-trivial one-loop graph
is the vertex graph, \fig{6}. Its value is\footnote{Our
$\gamma$-matrices are those of Bjorken and Drell\cite{23}, except that 
our wave
functions satisfy the normalization conditions $\bar u u=2m$, $\bar vv=-2m$.}
$$
\eqalign{
   {ig^2\over 16\pi^4} & (2\pi\mu)^{2 \epsilon }
\cr
   & \int\d{^{4-2 \epsilon}k}
   \, {\bar v(p_B)\gamma^\nu(-\st p_B+\st k+M) \gamma^\mu (\st p_A+\st k+M)
       \gamma^\lambda u(p_A)N_{\lambda\nu}(k) 
   \over 
       \left[-(p_B-k)^2+M^2-i\epsilon\right] (m^2-k^2-i\epsilon)
       \left[-(p_A+k)^2+M^2-i\epsilon\right]
    },
}
\eqno(25)
$$
where the numerator of the gluon propagator is
$$
N_{\lambda\nu}(k) = g_{\lambda\nu} -
    k_\lambda k_\nu(1-\xi)/\left(k^2-\xi m^2+i\epsilon\right).
\eqno(26)
$$                       
The possible reduced graphs are exactly the same as in $\phi^3$ theory and
are listed in \fig{3}.  

\begin{sloppypar}
Since our theory has a dimensionless coupling, the ultraviolet region,
of large $k$, contributes to the leading power of $Q$, 
just as in $\phi^3$ theory in six
dimensions. However, unlike $\phi^3$ theory in either dimension, the other
three regions also give leading contributions. This happens because the
numerator factor in \er(25) is $O(Q^2)$ in all three regions;
we know from our analysis of
$\phi^3$ theory that the graph would otherwise be of order $1/Q^2$. We must
first understand exactly how this factor of $Q^2$ arises, since a systematic
treatment of such enhancements by numerators is the key  to a complete
treatment of the form factor to all orders of perturbation theory. The mere
fact that the reduced graphs \fig{3}(b), (c) and (d) are leading twist
means that a simple factorization like \er(17) or \er(21) cannot
hold.
\end{sloppypar}

\subsec {Method of Grammer and Yennie}

Consider the region corresponding to the reduced graph of \fig{3}(b). This
is where $k^\mu$ is collinear to $p^\mu_A$ --- see \eq(11).
In this region the gluon is moving slowly relative
to the quark, $p^\mu_A$, and we may regard the gluon and the virtual quark
$p_A+k$ as being given a large boost from the center-of-mass frame. Thus
the term in
$$
(\st p_A+\st k+M)\gamma^\lambda u(p_A),
$$
with $\lambda = +$ is by far the largest.

It follows that the sum over $\nu$ in \er(25) is dominated by the term
with $\nu=-$. (Remember that in light-cone coordinates the metric is
non-diagonal: $g_{+-}=g_{-+}=1$, $g_{++}=g_{--}=0$.)

Note that we cannot say that the sum over $\lambda$ is dominated by
$\lambda=+$, because of the $k_\lambda k_\nu$ terms in the gluon
propagator. (Even if we set $\xi=1$, so that we used Feynman gauge, such
terms would arise when we consider graphs with vacuum polarization
corrections for the gluon).

A more general argument giving the same answer can be made by treating all
the $\gamma$-matrices as order 1. We wish to see how the numerator terms
with large components, viz.~$p^+_A$, $k^+$, $p^-_B$ contribute. To
do this, we anticommute all $\gamma^-$'s to the left and all $\gamma^+$'s
to the right and use ${\gamma^+}^2 = {\gamma^-}^2 = 0$. 
Then we use the mass-shell conditions
$$
\eqalign{
(\st p_A-M)u(p_A)&=0,\cr
\bar v(p_B)(\st p_B+M)&=0.}\eqno(27)
$$
Since $k^+$ and $p^+_A$ always multiply a $\gamma^-$ in the numerator and
since $p^-_B$ always multiplies a $\gamma^+$, we find that the large terms
only arise from anticommuting a $\gamma^-$ with the $\gamma^\lambda$ or a
$\gamma^+$ with the $\gamma^\nu$.

Let us write the numerator as
$$
\gamma^\nu(-\st p_B+\st k+M) \gamma^\mu
(\st p_A+\st k+m)\gamma^\lambda N_{\lambda\nu}=
B^\nu \gamma^\mu A_\nu,\eqno(28)
$$
with $B^\nu=\gamma^\nu(-\st p_B+\st k +M)$. To simplify this, we use a beautiful
trick formalized by Grammer and Yennie\cite{24}. It starts by making
the following string of approximations
$$
\eqalign{
B^\nu A_\nu &\simeq B^-A^+ \cr
&={1\over k^+}B^-k^+A^+\cr
&\simeq{1\over k^+}B^\nu k_\nu A^+\cr
&=k\cdot B{A\cdot u_B\over k\cdot u_B}.}\eqno(29)
$$
Here we rely on the facts that $+$ components of $A^\nu$ and $k^\nu$
are their largest and that the $-$ component of $B^\nu$ is not much smaller
than its other components. To put the result in covariant form, we have
defined $u^\mu_B$ to be a light-like vector with $u^-_B=1$, $u^+_B=u^T_B=0$.

We now have a factor $k_\nu$ times the lower vertex. This is a standard
situation where Ward identities can be used. In the present case the result
is easy to derive:
$$
\eqalign{
\bar v(p_B)k_\nu B^\nu&=\bar v\st k(-\st p_B+\st k+M)\cr
=&\bar v\left[(-\st p_B+\st k-M)+(\st p_B+M)\right](-\st p_B+\st k+M)\cr
=&\bar v\left[(p_B-k)^2-M^2\right],}\eqno(30)
$$
where we used the on-shell condition \er(27) for the wave function.
The factor $(p_B-k)^2-M^2$ cancels the antiquark propagator and we find
that if $k^\mu$ is restricted to be collinear to $p^\mu_A$, then
$$
\Gamma \simeq {-ig^2\over 16\pi^4}\ \int_{{\scriptstyle k\ {\rm collinear}
\atop \scriptstyle {\rm to\ }p_A}}\ \d{^4k}
{\bar v(p_B)u^\nu_B \gamma^\mu (\st p_A+\st k+M) \gamma^\lambda u(p_A)
\over
  k\cdot u_B\left(m^2-k^2\right) \left[M^2-\left(p_A+k\right)^2\right]}
N_{\lambda\nu}(k),
\eqno(31)
$$
with errors being smaller by a power of $1/Q$ (or $\lambda$, where
$\lambda$ is the small scale factor in \eq(11)).

The coupling of the gluon to the antiquark has become featureless; it is
in fact insensitive to the spin and energy of the antiquark. All the gluon
sees is the direction and charge of the antiquark. We do not need an
$i\epsilon$ prescription for the pole of $1/k\cdot u_B$ at $k^+=0$, for
$k^+$ is always large in the collinear region.

An exactly similar result holds for the opposite collinear region (\fig{3}(c))
$$
\Gamma\sim {-ig^2\over 16\pi^4} 
\int_{{\scriptstyle k\ {\rm collinear} \atop \scriptstyle {\rm to\ }p_B}} 
\ \d{^4k}
{   \bar v(p_B)\, \gamma^\nu ( -\st p_B+\st k+M) 
     \gamma^\mu u^\lambda_A u(p_A)
\over 
 [M^2 - (p_B-k)^2]
 (m^2-k^2-i\epsilon)
 (u_A \cdot k)
}
N_{\lambda\nu}(k).
\eqno(32)
$$

Furthermore a slightly different result holds if $k^\mu$ is in the soft region,
symbolized by the reduced graph \fig{3}(d):
$$
\Gamma \sim {ig^2\over 16\pi^4}\int_{\rm soft} \d{^4k}
{\bar v(p_B)u^\nu_B
    \gamma^\mu u^\lambda_A u(p_A)N_{\lambda\nu}(k)
\over
   \left(u_B\cdot k-i\epsilon\right)
   \left(m^2-k^2-i\epsilon\right)
   \left(u_A\cdot k+i\epsilon\right)
}.\eqno(33)
$$
The only subtlety in the derivation of this equation is that we must assume
that all components of $k$ are comparable (or at least that $|k^+/k^-|$,
$|k^-/k^+|\ll Q^2/M^2$, $|k^+k^-|\gsim k^2_T$). However, there is a leading
contribution when $k^+$ and $k^-$ are of order $\lambda^2Q$ and $k_T$ is
of order $\lambda Q$, with $\lambda$ a small quantity. This is the
Glauber region\cite{25}, and there none of 
the approximations \er(31) to \er(33) is valid.  
Now, in the Glauber region
$m^2-k^2\sim m^2+k^2_T$, independently of $k^+$ and $k^-$.  So
we can get out
of this region by deforming the contours of integration over $k^+$ and $k^-$
away from the poles in the quark and antiquark propagators to where at least one
of \er(31) to \er(33) is valid. To indicate the direction of
deformation, we introduced the $i\epsilon$'s with the $u_A\cdot k$ and
$u_B\cdot k$ denominators.

Note that there are overlap regions for $k$ where two or more of the
approximations \er(31) to \er(33) are simultaneously valid.
We will see shortly how to avoid the double counting that this could
give in the factorization theorem.

The physics of the general case is visible from the one-loop case. The
simplification in going from \eq(25) to any of \er(31)
to \er(33) is to replace one or both quark lines by an eikonal
approximation.  That is, 
the quark is replaced by a source of the appropriate
charge that exists along a light-like line in either the $+$ or $-$
direction, and recoil of the approximated quark is neglected. What is
happening is that there is a large relative rapidity between the gluon
and the approximated quark. The gluon only sees a Lorentz-contracted
object of a certain charge moving at the speed of light in a certain
direction. On the other hand the quark only sees the gluon for only a
short time in the quark's rest frame immediately before the
annihilation. 

\subsec {Leading regions for general graph}

The manipulations in the preceding sections have succeeded in simplifying
the integrand of \er(25) in the regions where $k^\mu$ remains close to
mass-shell as $Q$ goes to infinity. We will use these results, and their
generalization to higher order to construct a useful factorized form for
the complete form factor.

\deffigB{7}{%
   \centering
   \includegraphics[scale=0.95]{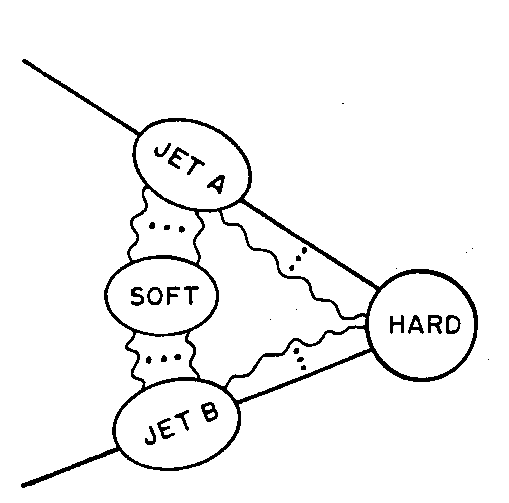}
}{Leading regions for quark form factor.  }

The first step (in the general case) is to see that, for a general graph
$\Gamma$, all the regions that give leading-twist contributions have\cite{11} 
the
form of \fig{7}. Each region is specified by dividing the graph into four
subgraphs, designated ``jet-A'', ``jet-B'', ``soft'', and ``hard''.
In the subgraphs jet-A and jet-B, the momenta of the internal lines satisfy
$|k^+|\gg |k^-|$ and $|k^-|\gg |k^+|$ respectively. A quark line from jet-A
and an antiquark from jet-B enter the hard subgraph, together with arbitrarily
many gluons.  The hard subgraph also includes the vertex for the
current.  
The momenta of the internal lines of the hard subgraph satisfy
$|k^2|\gg M^2$. The soft subgraph consists of lines all of whose momentum
components are much less than $Q$. The external lines of the soft subgraph
are all gluons and attach to one or other of the jet subgraphs.  Some
regions have no soft subgraph.  

Each of the jet subgraphs and the hard subgraph is connected. The soft
subgraph, if present, may consist of more than one connected component;
but each of its components must be joined to both jet-A and jet-B.

Note that within the jet and soft subgraphs there may be loops
with large ultra-violet momenta. These make up vertices of the reduced graphs
(as in \fig{1}). These reduced vertices are of the same form as the ordinary
vertices of the theory if the corresponding regions of momentum space are
to give leading twist contributions.

We must be more precise about the regions represented by \fig{7}.
The momenta satisfy the following requirements.
The value of $|k^+/k^-|$ for the momentum of a soft line must be
much less than the value of this ratio for the lines in the jet-$A$
subgraph 
and must be much greater than the ratio for the lines in the jet-$B$
subgraph.  This ensures that the Grammer-Yennie approximation is
applicable to the coupling of the soft lines to the collinear lines.
The momenta in the hard subgraph must have virtualities that
are much greater than for the momenta in the soft and collinear
subgraphs.  This ensures that the Grammer-Yennie approximation applies
to the coupling of collinear gluons to the hard part.

\begin{sloppypar}

The next step in the proof is to use the same approximations of the
Grammer-Yennie type that we used for the one-loop graph. After use of
Ward identities, this will give a factorization. Then we will write the
factors in terms of matrix elements of certain operators.  In general, a
given part of the space of loop momenta may be in the intersection of
several different regions of the form of \fig{7}.  We will have to make
an arbitrary choice of which region to use.  The resulting factorization
will involve graphs with momenta restricted to certain regions of
momentum space. We will convert this intermediate factorization to a
more useful factorization by showing that operator formulae representing
the second factorization can be converted to the same intermediate
factorization. 

\end{sloppypar}

The resulting factorization will still not be in a form that allows
perturbative calculations without large logarithms.  But it will enable
us to derive a differential equation for the $Q$-dependence of the form
factor. The solution of this equation will be our ultimate result, and
will have all the logarithms separated out. 

\subsec {Factorization}

\deffigA{8}{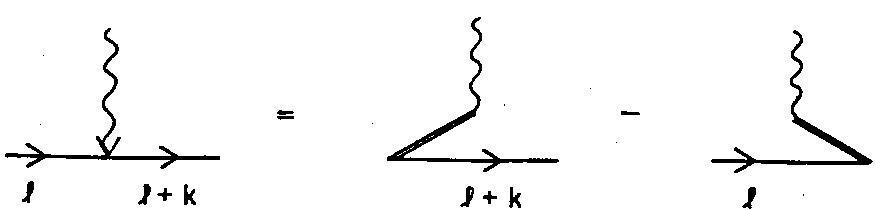}{Grammer-Yennie approximation and elementary Ward
identity.  }

When a gluon collinear to $A$ attaches to the hard subgraph, we use
approximation \er(29) where now $B^\nu$ denotes the elementary 
vertex where the collinear gluon attaches to the hard
subgraph. We then use a Ward identity of the sort illustrated in \fig{8}.
We let the vertex be $B^\nu=ig\gamma^\nu$ and the momenta of the quarks
be $l $ and $l+k$. Then
$$
\eqalign{
   {u^\nu_B k_{\nu'} \over k \cdot u_B}
   & \, \left[ {i(\st k+\st l +M) \over (k+l)^2-M^2} \right]
   ig\gamma^{\nu'}
   \left[ {i(\st l +M) \over l^2-M^2} \right]
\cr
   & \qquad \qquad = {i g u^\nu_B \over k\cdot u_B}  
      \left[ {i(\st k+\st l +M) \over (k+l)^2-M^2} \right]
      \left[ \st k +\st l - M - (\st l-M) \right]
      \left[ {i(\st l+M) \over l^2-M^2} \right]
\cr
   & \qquad \qquad = i g u^\nu_B
      \left\{ 
           {i \over k\cdot u_B} 
             \left[ {i(\st l+M) \over l^2-M^2} \right]
          -{i\over k\cdot u_B}
             \left[ {i(\st k+\st l +M) \over (k+l)^2-M^2}\right]
       \right\}.
}
\eqno(34)
$$
On the right of \fig{8} we use the double line to denote the eikonal
propagator $i/k\cdot u_B$. When we sum over all ways of attaching the jet
gluons to the hard subgraph, there is a whole set of cancellations and the
effect is to take the gluons to the $A$ side of the hard subgraph, as depicted
in \fig{9}. In obtaining this we have used \eq(34) repeatedly.
Each gluon has a factor $i/k\cdot u_B$. Then, we use identities like
$$
\left[ {i\over k_1\cdot u_B} \right] \left[ {i\over k_2\cdot u_B} \right]
=   \left[ {i\over (k_1+k_2)\cdot u_B} \right] 
    \left[ {i\over k_1\cdot u_B} \right]
   \, + \, 
     \left[ {i\over (k_1+k_2)\cdot u_B} \right] 
     \left[ {i\over k_2\cdot u_B} \right]
\eqno(35)
$$
to write the gluon attachments as if they are to a single line with an eikonal
propagator. Physically, \fig{9} is telling us that the gluons collinear
to $A$ only see the charge and direction of the antiquark.

\deffigB{9}{%
   \centering
   \includegraphics{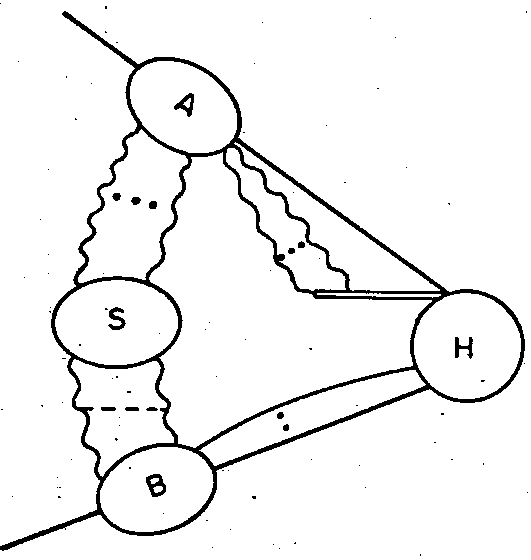}
   \vspace*{1cm} % to force fullpage, as in original
}{Result of applying Ward identities to attachments of gluons
from jet $A$ to the ultraviolet subgraph in \fig{7}.  }

Similar arguments applied to the attachments of the gluons collinear to
$B$ to the hard part and to the attachments of the soft gluons to the jets
give \fig{10}. Diagrammatically, \fig{10} represents a factorization, but
with the momenta in the subgraphs restricted to particular regions.  Notice
that since the momenta in the soft part are much less than the momenta in
the hard part, we ignore the dependence of the hard part on the loop
momenta that couple the soft to the hard part.

\deffigB{10}{%
   \centering
   \includegraphics[scale=0.95]{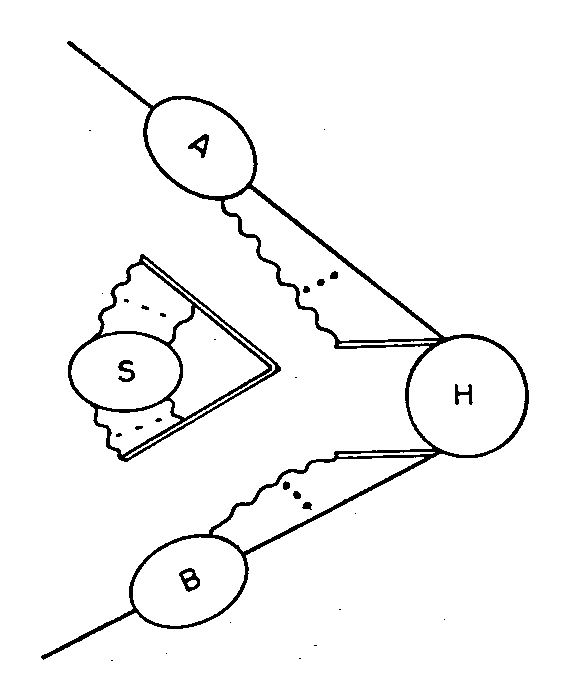}
}{Basic factorization of form factor.  }

Our aim now is to construct a formula 
that exhibits the factorization of \fig{10}, that has explicit
operator definitions of the factors, and that has no restrictions on the
momenta of the lines in the Feynman graphs.
First let us observe that, for example, the Feynman rules for both
the jet-A subgraph and the eikonal line attached to it can be derived from the
following matrix element of the quark field with a path-ordered exponential
of the gluon field:
$$
\left< 0\left|T\exp\left[-ig \mu^\epsilon 
\int^\infty_0 \d{z} u_B\cdot A(-u_B z)\right]
q(0)\right|p_A\right>.\eqno(36)
$$

Now consider the following quantity:
$$
\eqalign{
    J_A \left( \frac{p_A\cdot n^2}{n^2}; m,M,g,\mu \right) 
  \equiv {}&
    {\left<0|T\exp\left[ig \mu^\epsilon \int^\infty_0 \d{z}
      \,n\cdot A(n z)\right]q(0)| p_A \right>
   \over \left<0|T\exp\left[ig \mu^\epsilon \int^\infty_0 \d{z}\,n\cdot
     A(n z)\right]|0\right>
   }
\cr
   &\times  \hbox{UV renormalization factor}.}
\eqno(37)
$$
The numerator is the same as \er(36), except that we have replaced
$u^\mu_B$ by a space-like vector $n^\mu \equiv u_A^\mu - u_B^\mu$.
This has the effect of suppressing the contribution of momenta for which
$|k^-|\gg |k^+|$, i.e., momenta collinear to $B$. The need for the vector
$n^\mu$ to be a space-like
rather than time-like will appear later. The denominator in
\er(37) is necessary to cancel graphs, like \fig{11}, with 
eikonal self-interactions; these
do not appear in \fig{10}. Finally, since we have removed
all restrictions on loop momenta, there are ultra-violet divergences in
graphs like \fig{12}; these we define to be cancelled by renormalization
counterterms.

\deffigA{11}{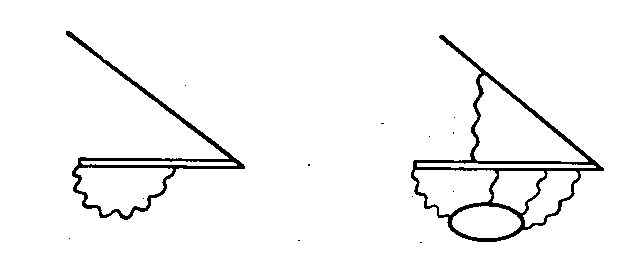}{Typical graphs with eikonal self interactions.  }

\deffigB{12}{%
   \centering
   \includegraphics[scale=0.95]{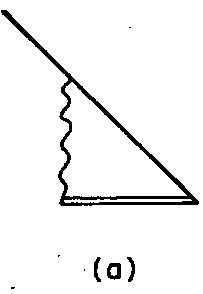}
   \includegraphics[scale=0.95]{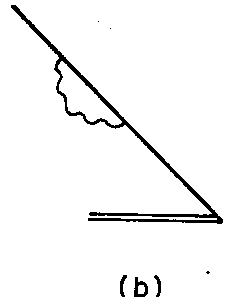}
}{One loop graphs for $J_A$.  }

The one-loop contributions to $J_A$ are given in \fig{12}, 
so that
$$
\eqalign{
   J_A = {}& z^{1/2}_2u(p_A)-{ig^2\over 16\pi^4}(2\pi\mu)^{2 \epsilon }
        \int \d{^{4-2 \epsilon}k}
         {\gamma^\mu \left(\st p_A+\st k+M\right) N_{\mu\nu}(k) n^\nu u(p_A)
         \over 
           \left(m^2-k^2\right) \left[M^2-(p_A+k)^2\right] (n\cdot k+i\epsilon)
         }
\cr
   &+ \hbox{UV counterterm} + O(\al^2).
}\eqno(38)
$$
This reproduces the contribution to \fig{6} of the region where $k^\mu$
is collinear to $A$. In \er(38), $z_2$ is the residue of
the pole of the quark propagator.

A jet-B factor may be defined similarly:
$$
\eqalign{
  J_B \left( \frac{{p_B\cdot n}^2}{ n^2}; M, m, g, \mu \right) 
  \equiv {}&
    {\left<0| \, T \, \bar q(0)\exp\left[ig \mu^\epsilon 
      \int^\infty_0 \d{z}\,n\cdot A(-n z)\right] |\bar p_B \right>
    \over \left<0|\, T \, \exp\left[ig \mu^\epsilon 
      \int^\infty_0 \d{z}\,n\cdot A(-n z)\right]|0\right>}
\cr
   &\times  \hbox{UV renormalization factor}.}
\eqno(39)
$$

We now apply the same argument to $J_A$ and $J_B$ as the one we applied
to obtain \fig{10} from \fig{7}. The result has the form
$$
\eqalign{
J_A &= (\hbox {Jet-$A$}) \times \hbox{soft factor}\times\hbox{hard factor},\cr
J_B &= (\hbox {Jet-$B$}) \times \hbox{soft factor}\times\hbox{hard factor},}
\eqno(40)
$$
where ``Jet-$A$'' and ``Jet-$B$'' are the same quantities as in \fig{10},
but the soft and hard factors are different. Hence the form factor can
be written
$$
F(Q) = J_A \left( \frac{{p_A \cdot n}^2}{ n^2} \right) 
\times J_B \left( \frac{{p_B \cdot n}^2}{ n^2} \right) 
\times \hbox {soft}\times \hbox {hard}+O(1/Q^2).
\eqno(41)
$$

We next recognize that the soft factor in \fig{10} has the Feynman
rules for
$$
\left<0 \left| T 
 \exp \left[ ig \mu^\epsilon 
\int^\infty_0 \d{z}\,u_A\cdot A\left(-z u^\nu_A\right) \right]
 \exp \left[-ig \mu^\epsilon 
\int^\infty_0 \d{z}\,u_B\cdot A\left(-z u^\mu_B\right) \right]
\right|0 \right>,\eqno(42)
$$
with the momenta restricted to the soft region.  Suppose we were to define
a soft factor as exactly \er(42) without any restrictions on loop momenta.
Then there would be divergences from regions where gluons become collinear
to $u_A$ or $u_B$.  These divergences are caused by the fact that $u^\mu_A$
and $u^\mu_B$ are light-like and effectively represent an incoming quark
and antiquark of infinitely high energy.  An example is given by the one
loop graph, \fig{13}:
$$
{ig^2\over (2\pi)^4}\int \d{^4k}
   {N_{+-}(k)
\over 
   (k^- + i\epsilon) (m^2 - k^2) ( k^+ - i\epsilon) }.
\eqno(43)
$$
The collinear divergences come from the regions where 
$k^+/k^-\to 0$ or $\infty$ with $k^+k^-$
fixed. They are
evidently artificial divergences.  The actual collinear regions
of the original form factor have already been taken into account by the
factors $J_A$ and $J_B$ in \eq(41), so we would also be guilty
of double-counting if we were to keep exactly 
\er(42) as our definition of the soft factor.  (There is also an ordinary
ultraviolet divergence from the region where $|k^\mu| \to \infty$; we will
deal with this separately.)  

\deffigA{13}{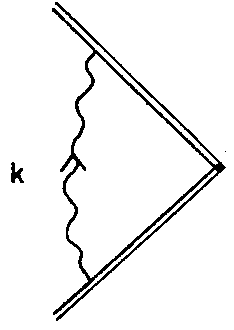}{One loop graph for \eq(42).  }

\deffigA{14}{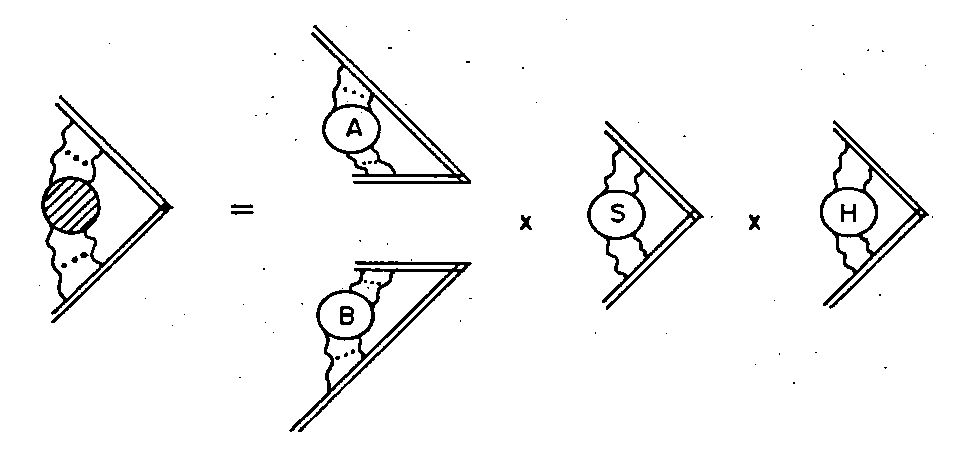}{Factorization for \eq(42).  }

What we must do is to observe that the argument that led to the
factorization of \fig{10} for the form factor can also be applied to
\er(42). The result is that the collinear parts factorize. This
is shown in \fig{14}, where the soft factor is identical to the soft
factor in \eq(41). Therefore we can write the original form factor as
$$
F=J_A (p_A\cdot n^2/n^2)  J_B (p_B\cdot n^2/n^2)
S(M,m,g,\mu)\times {\rm hard}+O(1/Q^2),\eqno(44)
$$
where $S$ is the quantity \er(42) divided by its collinear divergences:
$$
\eqalign{
S & \equiv {} \left<0\left|T\exp\left[ig \mu^\epsilon 
\int^\infty_0 \d{z}\,u_A\cdot
A\left(-z u^\mu_A\right)\right]\exp\left[-ig \mu^\epsilon 
\int^\infty_0 \d{z}\,u_B\cdot A\left(
-z u^\mu_B\right)\right]\right|0\right>\cr
&\times \left\{\left<0\left|T\exp\left[ig \mu^\epsilon 
\int^\infty_0 \d{z}\,u_A\cdot
A\left(-z u_A\right)\right]\exp\left[ig \mu^\epsilon 
\int^\infty_0 \d{z}\,n\cdot A\left(
z n\right)\right]\right|0\right>\right\}^{-1}\cr
&\times \left\{\left<0\left|T\exp\left[ig \mu^\epsilon 
\int^\infty_0 \d{z}\,n\cdot
A\left(-z n\right)\right]\exp\left[-ig \mu^\epsilon 
\int^\infty_0 \d{z}\,u_B\cdot A\left(
-z u_B\right)\right]\right|0\right>\right\}^{-1}\cr
&\times\left<0\left|T\exp\left[ig \mu^\epsilon 
\int^\infty_0 \d{z}\,n\cdot
A\left(z n\right)\right]\right|0\right>
\left<0\left|T\exp\left[ig \mu^\epsilon 
\int^\infty_0 \d{z}\,n
\cdot A\left(-z n\right)\right]\right|0\right>.}\eqno(45)
$$
This is represented by \fig{15}, and the last line in the equation has as
its sole purpose the canceling of eikonal self-energies. In \er(45)
we implicitly define UV divergences to be cancelled by renormalization
counterterms.

\deffigB{15}{%
   \centering
   \includegraphics[scale=0.85]{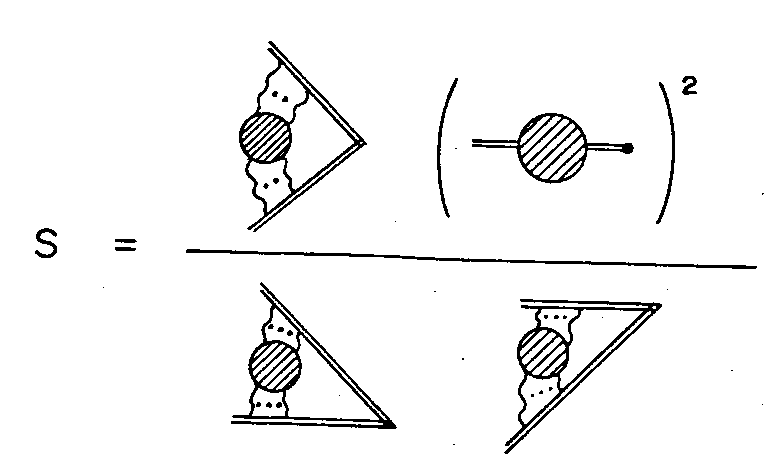}
}{Definition of soft factor.  }

We finally observe that when the form factor $F$ is divided by
$J_A J_B S$, all its collinear and soft regions have been cancelled.  
Hence we can define the hard factor by
$$
H(Q,g,\mu)\equiv \lim_{\scriptstyle M\to 0\atop \scriptstyle m\to 0}
F/(J_A J_B S).\eqno(46)
$$
Therefore we obtain the factorization
$$
\eqalign{
  F(&Q, m,M,g,\mu)=
\cr
    &J_A(-Q^2/4; M,m, g,\mu) \, J_B(-Q^2/4;M,m,g,\mu) \, S(m,M, g,\mu) \,
     H(Q;g,\mu) 
\cr
    &+O(1/Q^2).
}               
\eqno(47)
$$
Here $J_A$, $J_B$, $S$ and $H$ are defined by Eqs.~\er(37),
\er(39), \er(45) and \er(46). 

We cannot directly
use this equation to control the $Q$-dependence of $F$. However, the
dependence of $J_A$ and $J_B$ on $Q$ is through the vector $n^\mu$,
since $Q^2=-4p_A\cdot n^2/n^2=-4p_B\cdot n^2/n^2$. What we will do in
the next section is to compute the $Q$-dependence of $J_A$ and $J_B$ by
differentiating each with respect to $n^\mu$, holding the physical
momenta $p_A^\mu$ and $p_B^\mu$ fixed. The $Q$-dependence of $H$ will be
renormalization-group controlled just as for the $Q$-dependent factor in
$\phi^3$ theory. 

The reason for using a spacelike vector $n^\mu$ can now be explained. On the
various occasions that we factor out a soft region we needed to deform
integrals over gluon momenta away from the Glauber
region $|k^+ k^-| \ll k_T^2$,
just as in deriving 
\eq(33). For consistency all these deformations must
be in the same direction. In the case of a soft region for the jet factors
$J_A$ and $J_B$ this means that the $n\cdot k$ denominator must be
$(k^--k^++i\epsilon)$ to give the same direction of deformation as in
\er(33).

\sec {Evolution equation}

In this section we will derive equations for the $Q$-dependence of the form
factors, first in $(\phi^3)_6$ theory and then in a gauge theory.

\subsec {$(\phi^3)_6$}

The $Q$-dependence of the form factor \er(17) for $\phi^3$ theory
is under renormalization-group control --- see \eq(21) or
\er(22).  That is, after the factorization is obtained, we
may change $\mu$ to different values in the two factors to eliminate
all the large logarithms in their perturbation expansions.
So we need not bother to derive an explicit equation for
its $Q$-dependence. However, for the sake of the comparison with the
case of a gauge theory, we will nevertheless do so. 

From \eq(22) we have              
$$
\eqalign{
{\partial\ln F\over \partial\ln Q} 
   &=\hat \gamma\left(g(c_2Q)\right)+
  \beta\left(g(c_2Q)\right){\partial\over \partial g}\ln \Gamma\left(
  1/c_2,g(c_2Q)\right)+O(1/Q^2)
\cr
  &\equiv \tilde \gamma\left(g(c_2Q),c_2\right)+O(1/Q^2).
}
\eqno(48)
$$
The right-hand side of this equation can be expanded in powers of $g$. The
expansion contains no logarithms of $Q$ or of the masses, so it is valid
to approximate it by the first term or two, provided only that $g(Q)$ is
small.  Since $\phi^3$ theory is asymptotically free, such an approximation
is valid for all large enough values of $Q$. Indeed, given that\cite{26}
$$
{g^2(\mu)\over 64\pi^3} = 
{4/3\over \ln(\mu^2/\Lambda^2)}+O\left(\ln(\ln\mu)/
\ln^2\mu\right),\eqno(49)
$$
we find that 
$$
{\partial\ln F\over \partial\ln Q} = 
{-4/3 \over \ln\left(Q^2/\Lambda^2\right)
}+\cdots .\eqno(50)
$$
Here we have used the value of $\tilde \gamma$ that can be extracted from
the one-loop form factor \er(16):  
$$
\eqalign{
  \tilde\gamma&=Q{\partial\over\partial Q} \,\hbox{\eq(16)} +O(g^4)
\cr
  &= -g^2/(64\pi^3) +O(g^4).
}
\eqno(51)
$$
Note that the calculation of $\tilde \gamma$ from the vertex graphs 
may be done entirely in the massless theory. This results in
a saving of calculational effort.

Since the right-hand-side of \er(48)
is of order $g^2(Q)$ with no extra logarithms
of $Q$, the variation of the form factor for, say, a doubling of $Q$ is
small, of order $1/\ln(Q/\Lambda)$. This is comparable to the scaling
violations in deep-inelastic scattering. If we integrate over a wide range
of $Q$, say from a value $Q_0$ to a value of order $Q_0^2/\Lambda$, effects of
order unity arise.  If both $Q/\Lambda$ and $Q_0/\Lambda$ are large,
then
$$
F(Q)=F(Q_0)\left[{\ln(Q^2/\Lambda^2)\over \ln(Q^2_0/\Lambda^2)}\right]^{-2/3}
\left[1+O\left(\ln\ln(Q_0)/\ln(Q_0)\right)\right].\eqno(52)
$$
These results will be used as a standard of comparison when we have derived
the corresponding results for a gauge theory.

The accuracy of the results may be systematically improved by calculating
higher orders in perturbation theory.  

\subsec {Gauge theory jet factors}

In a gauge theory, the form factor satisfies the factorization equation
\er(47). The hard (or ultra-violet) factor $H$ has $Q$-dependence that
is renormalization group controlled, just as in $(\phi^3)_6$ theory. But
there is further $Q$-dependence in the jet factors. This comes essentially
from the possibility of emission of gluons of moderate transverse momentum
in a range of rapidity bounded by the two incoming particles. These gluons
are divided into what we may term left-movers and right-movers by the vector
$n_\mu$. The remaining contributions are put into the soft factor $S$ and
the hard factor $H$. The details of the 
cut-off on rapidity given by the vector $n^\mu$
are incorrect in the central region of finite center-of-mass rapidity. 
The errors are compensated
by the $Q$-independent factor $S$ if they correspond
to quanta of low transverse momentum. Quanta of large transverse momentum
are also included in all of the factors $J_A$, $J_B$ and $S$, but they are
incorrectly approximated; the hard factor $H$ was defined to cancel these
errors.

If we were to directly investigate the $Q$-dependence of the form factor,
we would have to find its dependence on $p^\mu_A$ and $p^\mu_B$. We would
have to trace the flow of these external momenta inside Feynman graphs,
and this would be a hard task.  (See Sen's work\cite{10,27} for details
on how to do this.)  
But if, instead, we examine the factors $J_A$
and $J_B$, we see that their $Q$-dependence comes from dependence on $p_A\cdot
n^2/n^2$ and $p_B\cdot n^2/n^2$ respectively. So it is sufficient to find
their dependence on $n^\mu$. This is very much easier because it involves
the result of differentiating in their definitions the
path-ordered exponentials of the gluon field, with respect to direction.
This gives a very simple result, as we will now see.

We have for $J_A$
$$
{\partial J_A\over \partial\ln Q}=\delta n^\nu{\partial J_A\over \partial
n^\nu},\eqno(53)
$$
where $\delta n^\mu$ is a backward-pointing time-like vector, normalized so that
$\delta n^2=-n^2$ and $n\cdot \delta n=0$. With our previous representation,
where $n^\mu=u^\mu_A-u^\mu_B$, we have $\delta n^\mu=-u^\mu_A-u^\mu_B$.
The light-like vectors $u^\mu_A$ and $u^\mu_B$ are defined by $u^+_A=u^-_B=1$,
$u^-_A=u^+_B=u^T_A=u^T_B=0$.

The $Q$-dependence of $J_B$ is obtained by the opposite variation of $n^\mu$:
$$
{\partial J_B\over \partial\ln Q}=-\delta n^\nu{\partial J_B\over
\partial n^\nu}.\eqno(54)
$$
Since this will result in an equation identical to that for the $Q$-dependence
of $J_A$, we will restrict our attention to $J_A$.

The result of differentiating the path-ordered exponential with respect
to its direction is
$$
\eqalign{
    \delta n^\nu{\partial\over \partial n^\nu} 
  & \exp \left[ig  \mu^\epsilon 
    \int^\infty_0 \d{z} \, n\cdot A(n z)\right]
\cr
  = {}& i g  \mu^\epsilon 
    \int^\infty_0 \d{z}
    \left[\delta n\cdot A + z\delta n\cdot \partial A\cdot n\right]
    \times \exp\left[ig \mu^\epsilon 
    \int^\infty_0 \d{z}\, n\cdot A\right],
}
\eqno(55)
$$
where we used the canonical equal-time commutation relations of the gluon
field to commute $A^0$ with $A^3$ and $\partial A^3/\partial t$ with $A^3$.
The resulting Feynman rules for $\partial J_A/\partial\ln Q$ are exhibited
in \fig{16} and \fig{17}.  The version shown in the second line of
\fig{16} comes from the second derivation which we now give.

%*************************************************
%\renewcommand\textfraction{0}%      Default: 0.2  Force Fig 16 to fit
%*************************************************
\deffigA{16}{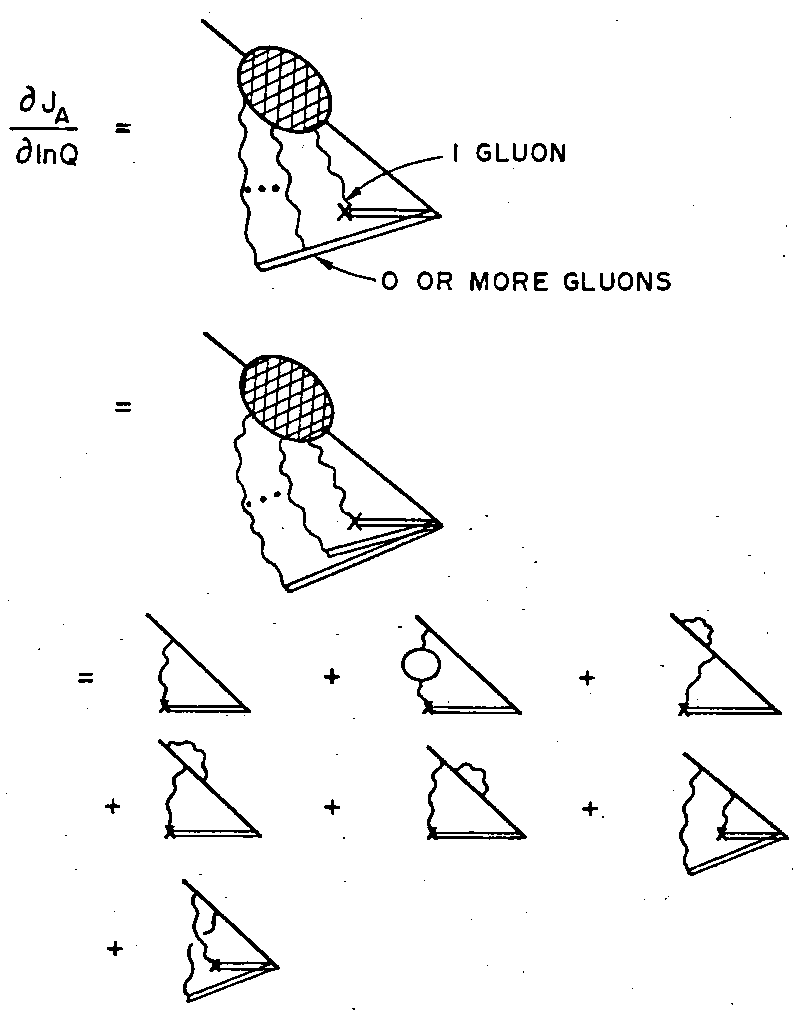}{Equation for $J_A$.  }
%  TYPESET EQN
%  For v.2: correspondence between diagrams and formulae reversed.
\deffigB{17}{%
   \begin{eqnarray*}
      \raisebox{-0.4\height}{\includegraphics[scale=0.9]{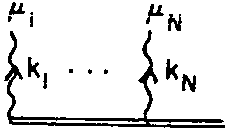}}~
   &=& 
      \left( ig n^{\mu_1} \frac{i}{ k_1 \cdot n + i\epsilon }\right)
      \dots
      \left( ig n^{\mu_N} 
             \frac{i}{ k_1 \cdot n + \dots k_N \cdot n + i\epsilon }
      \right)
   \\
      \raisebox{-0.4\height}{\includegraphics[scale=0.9]{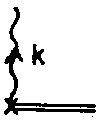}}~
   &=&
    \frac{ g \,  (\delta n \cdot k \, n^\mu  -  n \cdot k \, \delta n^\mu ) }
         { ( k\cdot n + i\epsilon )^2 }
   \end{eqnarray*}
}{Feynman rules for \fig{16}.  }

% For v.2: remove (a) on figure number
Suppose in a given Feynman graph (\fig{18}) for $J_A$ there are $N$ gluons
attaching to the eikonal line.  We have a factor
$$
{-g n^{\mu_1} \over (k_1\cdot n+i\epsilon)} 
\, {-g n^{\mu_2} \over (k_1\cdot n+k_2\cdot n+i\epsilon)}
\cdots 
{-g n^{\mu_N} \over (k_1\cdot n+k_2\cdot n+\ldots k_N\cdot n+i\epsilon ) }.
\eqno(56)
$$
Now let us sum over all graphs which are the same as the first one except
for having the $N$ gluons permuted. The result is to replace \er(56)
by
$$
{-g n^{\mu_1} \over (k_1\cdot n+i\epsilon)}
{-g n^{\mu_2} \over (k_2\cdot n+i\epsilon)}
\ldots
{-g n^{\mu_N} \over (k_N\cdot n+i\epsilon)},
\eqno(57)
$$
and is illustrated in
% For v.2
the left-hand side of
%
% For v.2: remove (b) on figure number
\fig{18}. Finally, we take the derivative given
on the right of \eq(53). This results  in a derivative for each
of the $N$ gluons, as symbolized in \fig{16},
each of the derivatives having the form
$$
{-g\Delta^\mu\over n\cdot k+i\epsilon } \equiv \delta n^\nu
{\partial\over \partial n^\nu}\left({-g n^\mu\over n\cdot k+i\epsilon}\right)=
{g\over (k\cdot n+i\epsilon)^2}
(\delta n\cdot k \, n^\mu - n\cdot k \, \delta n^\mu),
\eqno(58)
$$
as stated in the Feynman rules, \fig{17}.

\deffigB{18}{%
   \vspace*{10mm} % To force page size like original
   \centering
   \includegraphics{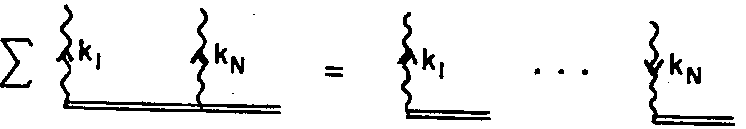}
}{Sum over attachments to eikonal line gives \eq(57).  }

To derive an evolution equation we now use the same style of argument that
we used to derive our first factorization \er(47). We can again divide
momenta into collinear, soft and hard. (The collinear momenta here are only
those collinear to $A$.)  The simplification that now occurs is that a 
collinear
momentum cannot enter the differentiated vertex.  This is easy to see, since
from \er(58) we find a factor
$$
A\cdot\Delta = {2 k^+A^- - 2 k^-A^+ \over k^- - k^+},
\eqno(59)
$$
which is suppressed by a large factor compared with 
$$
A\cdot n =A^--A^+\eqno(60)
$$
from a regular eikonal vertex.  Here $k^\mu$ and $A^\mu$ are vectors
collinear to $p_A^\mu$.
%*************************************************
%\renewcommand\textfraction{0.2}%      Default: 0.2 Undo earlier.
%*************************************************

\deffigA{19}{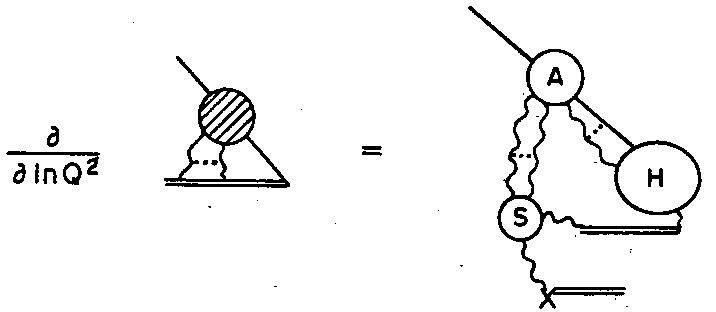}{Leading regions for $J_A$.  }

Therefore, to leading twist, only soft and hard momenta attach to the
differentiated vertex.  The result is \fig{19}, which is analogous to 
\fig{7}
for the form factor.

\deffigB{20}{%
   \centering
   \def\tmpscale{0.9}
   \includegraphics[scale=\tmpscale]{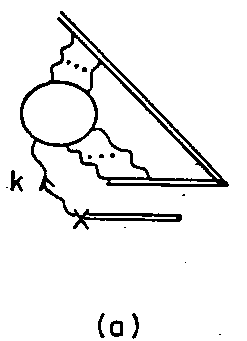}
   \includegraphics[scale=\tmpscale]{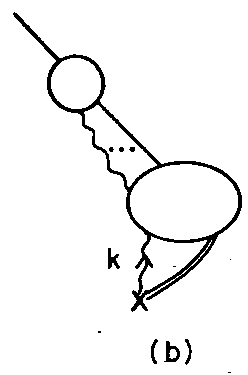}
}{Factor obtained from \fig{19} when $k$ is (a) soft, or (b) hard.}

Whenever $k^\mu$ is soft we can use the Grammer-Yennie method to obtain
the factor shown in \fig{20}(a). Whenever $k^\mu$ is hard its line disappears
into a hard subgraph; then we apply the derivation of \fig{9} to \fig{20}(b).
This gives
$$
{\partial J_A\over \partial\ln Q}=J_A\times \left[{\rm soft}+{\rm hard}\right]
+\hbox{higher twist}.\eqno(61)
$$

\subsec {Operator form of jet evolution}

It will be convenient to rewrite the soft factor in \er(61) so that
it has an explicit definition in terms of a matrix element of an operator,
just as we rewrote the jet and soft factors in \fig{10} to give
\eq(47). The same method of argument as for \fig{10} gives
$$
{\partial\ln J_A \over \partial\ln Q} = 
{1\over 2}K\left(m,M,\mu,g(\mu)\right) +{\rm hard}.
\eqno(62)
$$
We have defined the soft term $K$ to have a factor $1/2$, since it
will be multiplied
by 2 when we write the evolution equation for the form factor $F$. Furthermore,
this equation will have an extra hard term coming from the factor $H$ in
\eq(47), so we do not bother to name the hard term in \er(62).

\deffigA{21}{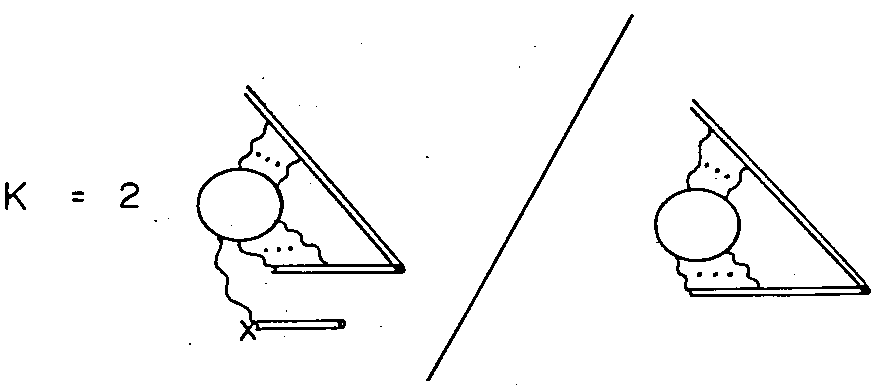}{Definition of $K$.  }

The definition of the quantity $K$ is given in \fig{21}. It can be written
as
$$
\eqalign{
   K={} 2
     &\left<0\left| T \exp\left[ig \mu^\epsilon 
                  \int^\infty_0 \d{z}\, u_A\cdot 
                          A\left(-z u_A\right) \right]
           \, \exp\left[ ig \mu^\epsilon 
                     \int^\infty_0 \d{z}\, n\cdot A(z n)\right]
      \right. \right.
\cr
   & \left. \left.
     \qquad ig \mu^\epsilon 
               \int^\infty_0 \d{z}
     \left[ \delta n\cdot A+z\, \delta n\cdot\partial(n\cdot A)\right]
        \right|0\right>
\cr
   &\left\{ \left<0\left|T \exp\left[ig \mu^\epsilon 
                \int^\infty_0 \d{z}
      \,u_A\cdot A(-z u_A)\right]
      \exp
         \left[ig \mu^\epsilon 
        \int^\infty_0 \d{z}\,n\cdot A(z n)\right] \right|0\right> 
   \right\}^{-1}
\cr
    &\hbox {with renormalization}.
}
\eqno(63)
$$
No restriction is placed on the internal momenta of $K$, so we must add
UV counterterms. Subdivergences in $K$ all correspond to the usual
renormalizations of the interaction, and the remaining divergence is an
overall divergence, so we must define it to be cancelled by an additive
counterterm:
$$
\eqalign{
   K (m, M, g,\mu) &= K_{\rm bare} + \hbox{counterterm}
\cr
   &=K_{\rm bare} + \delta K(g,\epsilon).
} \eqno(64)
$$
All soft contributions to $\partial\ln J_A/\partial\ln Q$ are contained
in $K$ together with some hard contributions.

\subsec {Gauge theory form factor}

We can now write an equation for the form factor
$$
\eqalignno{
  {\partial\ln F\over \partial\ln Q}&={\partial\ln J_A\over \partial\ln Q}+
  {\partial\ln J_B\over \partial\ln Q}+{\partial\ln H\over \partial\ln Q}+
  \hbox{higher twist}
\cr
  &=K(m,M,g,\mu) + 2 \times \hbox{hard} + 
  {\partial\ln H \over \partial\ln Q} +\hbox{higher twist}
\cr
  &\equiv K(m,M,g,\mu)+G(Q/\mu;g)+\hbox{higher twist}.
  &\eqalno(65)
}
$$
Since the dependence on masses has been separated from the dependence on
$Q$ on the right-hand side of this equation, we will have an effective
calculation of the large-$Q$ behavior of the form factor once we know the
renormalization-group equation for $K$.

This equation is easy to derive, since $K_{\rm bare}$ is 
renormalization group invariant: $(\mu
dK_{\rm bare}/d\mu=0)$.  So from \eq(64) we find that
$$
\eqalignno{
\mu{\d{K}\over \d{\mu}}&=\mu{\d{}\over \d{\mu}}\delta K\cr
&=\beta(g,\epsilon){\partial\delta K\over \partial g}\cr
&\equiv -\gamma_K(g).  &\eqalno(66)}
$$
The anomalous dimension $\gamma_K$ is finite at $\epsilon=0$, and if
we use minimal subtraction, 't Hooft's methods\cite{13} show that it is
independent of $\epsilon$. The anomalous dimension of the form factor is
zero, since the renormalized operator $[\jmu]$ has zero 
anomalous dimension:
$$
\mu {d\ln F\over d\mu} = 0.  \eqno(67)
$$
Thus \eq(65) gives the anomalous dimension of $G$:
$$
\eqalign{
\mu {d\over d\mu}G\left(Q/\mu;g(\mu)\right)&=-\lim_{m,M\to 0}\mu{dK\over
d\mu}\cr
&=\gamma_K\left(g(\mu)\right).}\eqno(68)
$$

We can now write the evolution equation in a form with no large logarithms:
$$
{\partial\ln F\over \partial\ln Q}=-\left[\int^{c_2Q}_\mu{ \d{\mu'}\over \mu'}
\gamma_K\left(g(\mu')\right)-G\left(1/c_2;g(c_2Q)\right)-
K\left(m,M,g(\mu),\mu\right)\right].\eqno(69)
$$
Here we displayed an overall factor $-1$, because the dominant
term is the integral over the positive 
one-loop value of $\gamma_K$.  

\sec {Interpretation}

The physical interpretation of \eq(69) and its derivation is as
follows:

When the total energy $Q$ is increased, the phase space available increases
for virtual quanta inside the form factor.  Let us consider the size of the
phase space split up into different ranges of transverse momentum.  At low
transverse momentum the range of phase-space is governed by the rapidity
between the quark and antiquark --- this gives the $K$ term.  At large
transverse momentum, of order $Q$, the quanta can only have finite
rapidity, but the range of transverse momentum increases with $Q$
--- this gives the $G$-term.  Finally, the intermediate range is filled in
by the anomalous-dimension term.

The precise form --- e.g., the $Q$-independence of $K$ --- and the detailed
derivation  rely on the fact that a particle (virtual or real) can only
probe details of another particle (e.g.\ the initial quark or antiquark)
if the relative rapidity is low. At large relative rapidity there is not
sufficient proper time to get a detailed picture. Indeed the only elementary
particle that can even interact at all across a large rapidity gap is the
spin-1 gluon. Then it just measures the total charge and the direction of
the probed particle.  A coherent sum over the detailed structure is needed
to give this result in perturbation theory, the result being formalized
in the Ward identities. It is the need to sum over a set of Feynman graphs
to get the physical answer that results in the technical complication of
our derivation.

The key to understanding the derivation is to ask what happens to quanta
inside the form factor when $Q$ is increased by boosting the quark and
antiquark in opposite directions. We first examine quanta with transverse
momenta much less than $Q$. The argument in the previous paragraph indicates
that the interactions of these quanta with other quanta of very different
rapidity, do not depend on the size of the rapidity gap. Thus the part of
the change in $F$ that comes from quanta of low transverse momenta can be
found by measuring the quanta that come into fill the interior of the increased
rapidity range.

To make this measurement we first measure the contents of the incoming quark
down to some finite rapidity, using the operator 
$$
\exp\left[ig \mu^\epsilon 
\int^\infty_0 \d{z}\,n\cdot A(n z)\right].
$$
Then we differentiate with respect to $n^\mu$,
to find the change caused by increasing the range of rapidity.

The remaining part of the variation of the form factor with $Q$ comes from
the short-distance regime of large transverse momentum.  This region is
well understood.

\sec {Solutions and calculations}

It is easy to solve \eq(69), with the result
$$
\eqalign{
  F(Q)={}&\hat F(m,M,g,\mu)\exp\Bigg\{-\int_\mu^{c_2Q}{\d{\mu'}\over \mu'}\left[
  \ln\left({c_2Q\over \mu'}\right)
  \gamma_K(g(\mu'))-G\left(1/c_2;g(\mu')\right)
  \right]
\cr
  &\qquad \qquad \qquad \qquad 
     +K\left(m,M,g(\mu),\mu\right)\ln(c_2Q/\mu)\Bigg\}
\cr
  &+O(1/Q^2),
}
\eqno(70)
$$
where $\hat F$ represents the effect of the initial condition for \eq(69),
and is determined by the form factor at low $Q$. 
If $\mu$ is chosen to be of order the masses then this is a form in which
no large logarithms appear in the coefficients of perturbation expansions;
the logarithms in the unimproved perturbation series for $F$ are either
explicitly in the exponent in \er(70) or are generated by the integration
over $\mu'$.

The leading logarithmic approximation to $F(Q)$ is obtained by using the
result (see later) that 
$\gamma_K=g^2/(2\pi^2)+O(g^4)$ and by writing $g(\mu')$
in terms of the running coupling, $g(\mu)$, at a fixed scale.  Then the
highest power of a logarithm of $Q$ is obtained from the lowest order term
in the exponent.  The result is
$$
F=\exp\left[-{g^2 \over 16\pi^2}\ln^2Q^2+\hbox{non-leading logs} \right].
\eqno(71)
$$

In QCD, formulae like \er(70) can be derived for a number of important
cases, such as the transverse-momentum distribution of the Drell-Yan 
process\cite{4,5}.
Since QCD is asymptotically free, we can do an effective calculation from
low orders of perturbation theory if $Q$ is large. Non-leading logarithmic
corrections are thereby tamed.

\subsec {Calculations}

We now consider how to calculate the quantities $K$, $G$ and, particularly,
$\gamma_K$ that appear in the exponent in \er(70) and on the right-hand
side of the evolution equation \er(69).  One method is to start from the
Feynman graphs for the form factor.  Then the results of these calculations
are compared with the general form \eq(70).  This determines $K$ and $G$
(and hence $\gamma_K$) except for an ambiguity of adding some function of
$g$ to $K$ and subtracting the same function from $G$.  This ambiguity is
the same as the renormalization-scheme ambiguity for $K$ and $G$, and as
such does not directly pertain to physics:  the physics resides in the
dependence of the functions $K$ and $G$ on their other arguments.
Especially at higher-order this procedure is rather lengthy. 

Now the terms
in the exponent with the largest powers of logarithms are the most
important --- in particular the $\gamma_K$ term. So a short-cut can be made
by calculating $K$. At two-loop order there is only one easy graph for $K$
--- a
vacuum polarization correction --- but for the form factor there are five
harder graphs.  The anomalous dimension $\gamma_K$ is obtained from $K$ and
in particular from its ultra-violet divergence.  

In addition, the form \er(70) implies many relations between the
logarithms of $Q$ in different orders. These are not manifest in a
direct calculation.  However their validity provides nontrivial tests of
calculations.  
Nevertheless all the coefficients can be obtained
by a direct calculation.  For example, from an evaluation of the
one-loop vertex, \eq(25), one can show that 
$$
   {\partial\ln F\over \partial\ln Q}
=  -\alpi \ln\left(\frac{-Q^2 }{ m^2} \right) + O( \al^2 )
   + \hbox{higher twist}.
\eqno(72)
$$
This is evidently of the form of the evolution equation, \eq(65).
It implies that 
$$
\eqalign{
   K &= \alpi \left[\ln(m^2) + C \right] +  O( \al^2 ),
\cr
   G &= - \alpi \left[ \ln(-Q^2) + C \right] + O(\al^2 ),
} 
\eqno(73)
$$
where $C$ is a constant.  The value of $C$ is not a priori
fixed, and a change of the constant corresponds to a change of
renormalization scheme for $K$.
We choose to resolve the ambiguity by using
\MSbar\ renormalization applied to a direct calculation of $K$ from
its Feynman rules.  Given that to 1-loop
$$
\eqalign{
   K&= {-ig^2 \over 8\pi^4}(2\pi\mu)^{2 \epsilon } \int \d{^{4-2 \epsilon}k}
   {(u_A\cdot n \, k\cdot \delta n \, - k\cdot n \, \delta n\cdot u_A)\over
   (m^2-k^2-i\epsilon)(u_A\cdot k+i\epsilon)(n\cdot k+i\epsilon)^2}+ 
   \hbox {counterterm}
\cr
   &= {-ig^2 \over 8\pi^4}(2\pi\mu)^{2 \epsilon } \int \d{^{4-2 \epsilon}k}
   {1\over (m^2-k^2-i\epsilon) (-k^z +i\epsilon)^2}+{\rm counterterm},
}
\eqno(74)
$$
it is fairly easy to perform the integrals.  The result, with \MSbar
\ renormalization, is that
$$
   K= \alpi \ln\left(m^2/\mu^2\right) + O (\al^2 ),
\eqno(75)
$$
from which follows
$$
   \gamma_K = 2\alpi + O( \al^2 ). 
\eqno(76)
$$

It is left as an exercise for the reader to show that 
the sole 2-loop graph gives the $O(\al^2)$ term in $\gamma_K$:
$$
\gamma_K= 2 \alpi -{10\over 9} \left( \alpi \right)^2
+O\left( \al^3 \right).\eqno(77)
$$

There is a lot of information in these results, even without the two-loop
result \er(77). For example, let us expand $\, \ln F$ in powers of
$t=\ln(-Q^2/ \mu^2)$
$$
\eqalign{
   \ln F = {}& \alpi \left(C_{12}t^2+C_{11}t+C_{10}\right)\cr
   &+\left( \alpi \right)^2\left(C_{24}t^4+C_{23}t^3+C_{22}t^2
   +C_{21}t+C_{20}\right)
\cr
   &+O(1/Q^2),
}
\eqno(78)
$$
where the coefficients may depend on $m$, $M$ and $\mu$, but not on $Q$.
The leading logarithm results imply that $C_{24}=0$.  Our formula \er(65)
implies considerably more. Now, from \er(78) we have
$$
\eqalign{
{\partial\ln F\over \partial \ln Q}={}& \alpi \left(4C_{12}t
+2C_{11}\right)\cr
&+\left( \alpi \right)^2\left(6C_{23}t^2+4C_{22}t+2C_{21}\right)\cr
&+\ldots.}\eqno(79)
$$
In order that $G$ in \eq(65) be independent of the masses $m$ and $M$,
$C_{12}$, $C_{23}$ and $C_{22}$ must be independent of $m$ and $M$ (and
hence of $\mu$).  Furthermore, once one puts in the one-loop values, the
requirement that $G$ satisfies its
renormalization group equation implies that
$$
C_{23} =-{1\over 36}. \eqno(80)
$$

Hence the 
new information for the form factor $F$
at 2 loops is 2 logarithms down from the leading logarithm, i.e.\ it is in
$C_{22}$ and the less leading coefficients, $C_{21}$ and $C_{20}$.  
The double logarithm coefficient $C_{22}$
is related to the two-loop term in $\gamma_K$, given
in \eq(77); this was the result of a relatively easy calculation.  Hence
$$
C_{22} = {5\over 36}.  \eqno(81)
$$
The remaining information, for which a full two-loop calculation of
the form factor is needed, is in the terms with one and no logarithms of
$Q$. These are three and four logarithms down from the leading $\ln^4Q$
term.

\subsec {Comparison with other work}

One can verify \er(77) and \er(80) 
from the calculation of Barbieri \etal\cite{28}. Note that 
in this calculation one must change renormalization prescription first.

Korthals-Altes and de
Rafael\cite{29} made a conjecture about an evolution equation for
the form factor we are discussing.  Their conjecture is
that $(Q \, \partial / \partial Q - \beta \, \partial /
\partial \al )\ln F$ is linear in $\ln Q$. 
It can be checked that their conjecture is implied by 
our \eq(69). However the converse
is not true: Their conjectured result does not imply our \eq(65)
with its specific dependences on masses.

A number of calculations of comparable quantities in QCD have been made.
As will be discussed below, generalizations of our formulae apply not only
to a simple quark form factor but notably also 
to the transverse momentum distribution in the Drell-Yan process and to 
two-hadron-inclusive production in $\epem$ annihilation.  (This last
includes the energy-energy correlation as a special case.)  The anomalous
dimension $\gamma_K$ is common to all these processes.  

The
electromagnetic form factor of a quark in massless QCD 
also satisfies our equation \er(65) or
\er(69), as shown by Sen\cite{10}. The coefficients are now 
$$
\eqalign{
   K&= \alpi C_F \left[ { 1 \over\epsilon} - \gamma + \ln (4\pi ) \right]
       +O\left( \al^2 \right),
\cr
   G&=- \alpi C_F\left[\ln\left(Q^2/ \mu^2\right)
       -{3\over 2}\right] + O\left( \al^2 \right),
\cr
   \gamma_K &= 2C_F \alpi  + 
       \left[ \left(\frac{67}{18}- \frac{\pi^2}{6}\right) C_F C_A
                 - {10\over 9} n_f C_F T_F \right] \left( \alpi \right)^2
       + O\left( \al^3 \right).
}\eqno(82)
$$
Here $n_f$ is the number of quark flavors, while 
$T_F$, $C_F$ and $C_A$ are the
usual group theory coefficients. ($T_F= \frac12$, 
$C_F=\frac43$ and $C_A=3$ for QCD, while $T_F= 1$, 
$C_F=1$ and $C_A=0$ for QED.)  Since
there are no masses, $K_{ \rm bare}$ is zero, and the renormalized $K$
equals its \MSbar\ counterterm.  The resulting pole, as displayed in
\eq(82), represents the infra-red divergence in $K$.
Those terms that appear in the abelian case
are given by our earlier calculations.  The only purely nonabelian term in
the order to which we work in \eq(82) is the two-loop $C_F C_A$ term.  We
have deduced its value from \cite{30,31,32}, as I will explain
later.

The calculation in QCD that can most directly be compared with \eq(82) is
by Gonsalves\cite{33} who has calculated precisely the quark form factor in
QCD, at two-loop order.  (The purpose of doing this is that one can use the
deduced value of $\gamma _K$ in other processes.)  
Gonsalves' results do not obey the correct evolution equation,
which should hold in QCD as well in QED.  
Note that his renormalization prescription differs in
detail from both \MSbar\ and MS.  The agreement between the other
calculations indicates that there must be an error in Gonsalves'
calculation.

\subsec {Infrared divergences in QCD}

Korchemskii and Radyushkin\cite{30,34} have studied the infrared divergences
of the electromagnetic form factor of a quark in QCD at large $Q$.
(Their ultimate aim\cite{35} is to study the full Sudakov
problem in QCD.)  Consequently their methods have much in common with the
work described in this paper.  Indeed their results are written in
terms of path-ordered exponentials that are similar to the ones used
in this article.  Moreover they have derived the necessary
generalization of the Grammer-Yennie method to the nonabelian case.

In an abelian theory, the infrared divergences (as the gluon mass
goes to zero) are rather simple: they form a factor which is the
exponential of the one loop infrared divergence.  In \eq(69),
for the $Q$-dependence of the form factor, all the infrared divergences
are in the term $K$.  Gluon self couplings in an abelian theory are
induced solely by quark loops, and these loops are suppressed when
the gluon momenta go to zero and the quark mass is nonzero.  Hence only the
one loop part of $\gamma _K$ is needed for the calculation of the infrared
divergences in the abelian theory.  

In QCD the infrared divergences are much more complicated, since
the gluon self couplings are not so suppressed at zero momentum.
Korchemskii and Radyushkin show that the infrared divergences form a factor:
$$
\msoft \equiv
 \langle 0| 
    T \bar P \exp [-i g \int_0^\infty \d{s}  \, p_A^\mu A_\mu (p_A s)]
    \, P \exp [i g \int_0^\infty \d{s}  \, p_B^\mu  A_\mu (p_B s)]
  | 0 \rangle _{\rm IR}.
\eqno(83)
$$
The subscript `IR' means that integrations are restricted to the infra-red
region.  When $Q^2 \equiv (p_A-p_B)^2$ gets large, 
the infra-red behavior is governed by the anomalous dimension of the cusp,
$\gcusp$.  The ability to do systematic perturbative calculations
at large $Q$ relies on the property, proved by Korchemskii and 
Radyushkin, that $\gcusp$ is linear in $\ln(Q)$ for large $Q$:
$$
 \gcusp = A(\al) \ln (Q/M) + B(\al) + O(1/Q^2).
\eqno(84)
$$
This linearity is implied by our results (if it is assumed that they
extend to QCD).  Indeed $A(\al)$ in \eq(84) is the same as
our $\gamma_K$.\footnote{Note that in going to the regime of infrared
divergences, it is necessary to compute the anomalous dimensions
in the effective low-energy theory that exhibits the decoupling of
massive quarks.}  The proof is simple:
$$
\eqalign{
   {\partial \over \partial \ln Q^2} \gcusp 
   &\equiv - {\partial \over \partial \ln Q^2} {\d{} \over \d{\ln \mu}} 
    \ln \msoft 
\cr
   &= - {\d{} \over \d{\ln \mu}} {\partial \over \partial \ln Q^2} 
    \ln \msoft
\cr
   &= - {\d{} \over \d{\ln \mu}} (K + G_{\rm KR}).
\cr
   &= - {\d{} \over \d{\ln \mu}} K 
\cr
   &= \gamma_K(\al).
}
\eqno(85)
$$
Here we have employed the factorization theorem for $\ln \msoft$
that is analogous to the one for the quark form factor.  The soft
term $K$ is the same as for the form factor.  But the hard term $G_{\rm
KR}$ is different.  Indeed, since it is a pure ultraviolet quantity, with
no scale dependence, it is zero.

This result enables us to obtain the non-abelian part of $\gamma_K$ at two
loop order; this is the term in \eq(82) that is proportional to $C_F C_A$.
Since Korchemskii and Radyushkin investigate the infrared divergences of
the form factor with massive quarks, they cannot calculate the $C_F T_F$
term.

\subsec {Other QCD calculations}

Kodaira and Trentadue\cite{32} considered the energy-energy correlation in
$\epem$ annihilation.  They worked with a different formalism for the
Sudakov form factor.  Their calculation was the first from which 
the nonabelian part of the two-loop value for $\gamma_K$ can be deduced.
They also agree with the abelian part, as calculated directly from $K$.

Davies and Stirling\cite{31} have calculated the Drell-Yan cross section at
order $\al^2$.  They deduce $\gamma _K$ and the equivalent of $K$ and $G$.
They confirm the value for $\gamma_K$ given in \eq(82).

\sec {Applications to QCD}

As has already been noted, there are many cases in QCD where something like a
Sudakov form factor enters.  
The most straightforward extension of the results in this article
is to transverse momentum distributions.  
The transverse-momentum $q_T$ is a third important scale for the cross-section,
in addition to the total energy $Q$ and the hadron mass-scale. Two logarithms
of $Q/q_T$ per loop are present in Feynman graphs.

In the case of two-particle inclusive cross-sections in $\epem$
annihilation, Collins and Soper\cite{4} derived an equation generalizing
Eqs.~\er(65) and \er(69). The same anomalous dimension
$\gamma_K$ makes its appearance. Technically the main difference between
\cite{4} and the treatment in the present article, aside from
having a non-abelian gauge group, was that there we used an axial gauge
$n\cdot A=0$ instead of Feynman gauge. This resulted in a nice
simplification. For example, in \eq(37), the line integral of
the gluon field is zero, so that in \cite{4} the definition of $J_A$
would have been 
$$
J_A\left(P_A\cdot n^2/n^2\right)=\left<0|q(0)|P_A\right>_{\rm axial\ gauge},
\eqno(86)
$$
with the dependence on $n^\mu$ now being a dependence on the choice of gauge.

My treatment of the Sudakov form factor in \cite{21} used
Coulomb gauge, which behaves for this purpose rather like the axial gauge.
In either gauge, explicit Feynman graph 
calculations are made more difficult than in covariant
gauge by the complicated form of the numerator of the gluon propagator. In
axial gauge we have
$$
N^{\rm axial}_{\mu\nu}=\left\{g_{\mu\nu}-{n_\mu k_\nu+k_\mu n_\nu\over n\cdot
k}+{k_\mu k_\nu n^2\over k\cdot n^2}\right\}_{\rm PV},\eqno(87)
$$
where `PV' denotes the principal-value prescription for the singularities
at $n\cdot k=0$. In Coulomb gauge, we have
$$
N^{\rm Coulomb}_{\mu\nu}=
g_{\mu\nu}-{(\delta n_\mu k_\nu+k_\mu\delta n_\nu) \, 
\delta n\cdot k\over -k^2\delta n^2+(\delta n\cdot k)^2}+
{k_\mu k_\nu\delta n^2\over -k^2\delta n^2+\delta n\cdot k^2},
\eqno(88)
$$
where $\delta n^\mu$ is the vector defined just below \eq(53).  
% Note that
% in the rest frame of $\delta n^\mu$ we have $-k^2\delta n^2+\delta n\cdot
% k^2=\vec k^2$.

\begin{sloppypar}
However, there are more fundamental disadvantages than calculational
complexity to use of these physical gauges.  It is very hard to define
higher order graphs in the axial gauge because of the need to multiply
principal values.  To overcome this, considerable complication in
the Feynman rules is necessary\cite{36}, and the simplicity of the
Ward identities is no longer clear.  There are also complications
in the Feynman rules in Coulomb gauge beyond 2-loop order\cite{37}.
In both cases, it is not clear that a complete and {\em correct}
all-orders derivation can be given easily.  
\end{sloppypar}

We expect corresponding results to the ones for the
energy-energy correlation to hold for the Drell-Yan process; they
have been formulated by Collins, Soper and Sterman\cite{5}. 
This work, because it entails a complete treatment of a factorization
theorem for a process at low transverse momentum, includes treatment of
intrinsic transverse momentum within QCD.  In other work of that period,
based on leading logarithmic formulations, intrinsic transverse momentum
tends to appear as an ad hoc phenomenological modification to the basic
formula for the cross section.

\begin{sloppypar}

Davies and Stirling\cite{6,31} have applied this formalism
phenomenologically.  Altarelli \etal\cite{7} have also performed
phenomenological calculations, but without the full treatment of the
intrinsic transverse momentum effects.

\end{sloppypar}

A further disadvantage to using the physical gauges 
appears\cite{1} when we try to derive results for the Drell-Yan cross-section.
The problem is that singularities in the numerators of the gluon propagators
\er(87) or \er(88) wreck the derivation of the form of the leading
regions.  Specifically, we need contour-deformation arguments generalizing
those which we summarized at the end of Sec.~5, and these are invalid in
an axial or Coulomb gauge. So Collins, Soper and Sterman\cite{1} were forced
to the use of a covariant gauge, at the price of some extra technicalities
in the proofs. At the same time, the proofs come out to be cleaner. It is
a generalization of 
the method of \cite{1} that is used in the present article.  

Another line of development comes from realizing that similar physical
phenomena to those in the Sudakov form factor occur inside of amplitudes
for scattering in the Regge region. Sen\cite{27} has produced very important
results in this area. He used Coulomb gauge.

\mainhead{ACKNOWLEDGEMENTS}

This work was supported in part by the U.S. Department of
Energy, Division of
High Energy Physics, contracts W-31-109-ENG-38 and DE-FG02-85ER-40235, and
also by the National Science Foundation, grants Phy-82-17853 and
Phy-85-07627,
supplemented by funds from the National Aeronautics and Space
Administration.  
I wish to thank the Institutes for Theoretical Physics at Santa Barbara and
Stony Brook 
for their hospitality during part of the preparation of this article.

\mainhead{REFERENCES}

\begin{enumerate}

\begin{sloppypar}

\defref{1} J.C.  Collins, D.E.  Soper and G.  Sterman,
    \np\rf{B261}{104}{85} and \rf{B308}{833}{88};
     G. Bodwin, \pr\rf{D31}{2616}{85} and \rf{D34}{3932}{86}.  These 
     papers give the fullest proofs of factorization in the case of the 
     Drell-Yan and other processes in hadron-hadron scattering.  See also
     Ref.~\ref{2} for inclusive $\epem$ cross sections, and Ref.~\ref{38}
     for the original papers.  A summary is given in the
     article by Collins, Soper and Sterman in this volume.
     % Factorization for Drell-Yan
\defref{2}  J.C.  Collins and G.  Sterman, \np\rf{B185}{172}{81}.
     %  Soft partons
\defref{3} V.  Sudakov, Zh.\ Eksp.\ Teor.\ Fiz.  \rf{30}{87}{56}; (Eng.\
    trans) Sov.\ Phys.\ JETP \rf{3}{65}{56}.
\defref{4} J.C.  Collins and D.E.  Soper, \np\rf{B193}{381}{81}, and
    \np\rf{B197}{446}{82}.  % small q_T energy^2 correlation.
\defref{5} J.C.  Collins, D.E.  Soper and G.  Sterman,
    \np\rf{B250}{199}{85}; J.C. Collins and D.E. Soper, ``Parton transverse
    momentum'', in ``Lepton Pair Production'' ed.\ J. Tran Thanh Van
    (Editions Fronti{\`e}res, Dreux, 1981).  
    % QT in DY
\defref{6} C.T.H.  Davies, B.R.  Webber and W.J.  Stirling,
    \np\rf{B256}{413}{85}.  % Drell-Yan q_T phenom.
\defref{7} G.  Altarelli, R.K.  Ellis, M.  Greco and G.  Martinelli,
    \np\rf{B246}{12}{84}.  % Drell-Yan q_T
\defref{8} Yu.L. Dokshitzer, D.I. Dyakonov and S.I. Troyan, Phys.\ Reports
    \rf{58}{269}{80}.
\defref{9} G. Parisi and R. Petronzio, \np\rf{B154}{427}{79}.
\defref{10} A. Sen, \pr\rf{D24}{3281}{81}.
   % Sudakov in QCD.
\defref{11} S.  Libby and G.  Sterman, \pr\rf{D18}{3252, 4737}{78}.
\defref{12} S.  Coleman and R.E.  Norton, Nuovo Cim.~\rf{28}{438}{65}.
\defref{13} G.  't Hooft, \np\rf{B61}{455}{73}.
\defref{14} W.A.  Bardeen, A.J.  Buras, D.W.  Duke and T.  Muta,
   \pr\rf{D18}{3998}{78}. 
\defref{15} M.  Creutz and L.-L.  Wang, \pr\rf{D10}{3749}{74}; S.-S.
   Shei, \pr\rf{D11}{164}{75}; P.  Menotti, \pr\rf{D11}{2828}{75}. 
\defref{16} See any good modern 
     textbook on field theory, e.g., G.  Itzykson and
     J.-B.  Zuber, ``Quantum Field Theory'' (McGraw-Hill, New York, 1980),
     or see J.C.  Collins, ``Renormalization'' (Cambridge University Press,
     Cambridge, 1984).
\defref{17} E.g.  D.  Gross in ``Methods in Field Theory'' (eds.  R.
    Balian and J.  Zinn-Justin) (North-Holland, Amsterdam, 1976), or
    Collins, Ref.~\ref{16}.
\defref{18} G. Grunberg, \pl\rf{95B}{70}{80} and \pr\rf{D29}{2315}{84};
     P.M. Stevenson, \pr\rf{D23}{2916}{81} and \np\rf{B203}{472}{82};
     D.W. Duke and R.G. Roberts, Phys.\ Reports \rf{120}{275}{85}.
     %  FAC, PMS etc. 
\defref{19} R.  Jackiw, Ann.\ Phys.~(N.Y.)~\rf{48}{292}{68}.
\defref{20} A.H.  Mueller, \pr\rf{D20}{2037}{79}.
\defref{21} J.C.  Collins \pr\rf{D22}{1478}{80}.
\defref{22} J.C.  Collins, Argonne preprint ANL-HEP-PR-84-36.
\defref{23} J.D.  Bjorken and S.D.  Drell, ``Relativistic Quantum Fields''
    (McGraw-Hill, New York, 1966).
\defref{24} G.  Grammer and D.  Yennie, \pr\rf{D8}{4332}{73}.
\defref{25} G.  Bodwin, S.J.  Brodsky and G.P.  Lepage,
    \prl\rf{47}{1799}{81}.
\defref{26} A.J.  Macfarlane and G.  Woo, \np\rf{B77}{91}{74}.
\defref{27} A.  Sen, \pr\rf{D27}{2997}{83} and \rf{D28}{860}{83}.
\defref{28} R.  Barbieri, J.A.  Mignaco and E.  Remiddi, Nuovo Cim.\
    \rf{11A}{824}{72}.
\defref{29} C.P.  Korthals-Altes and E.  de Rafael, \np\rf{B106}{237}{76}.
\defref{30} G.P. Korchemskii and A.V. Radyushkin, Yad.\ Fiz.\ 
   \rf{45}{1466}{87} [Eng.\ transl.: Sov.\ J.  Nucl.\ Phys.\
   \rf{45}{910}{87}].
\defref{31} C.T.H.  Davies and W.J.  Stirling, \np\rf{B244}{337}{84}.
     % Drell-Yan q_T theory
\defref{32} J. Kodaira and L. Trentadue, \pl\rf{112B}{66}{82}.
\defref{33} R.J.  Gonsalves, \pr\rf{D28}{1542}{83}.  
\defref{34} S.V.  Ivanov, G.P.  Korchemskii and A.V.
   Radyushkin, Yad.\ Fiz.  \rf{44}{230}{86} [Eng.\ transl: Sov.\ J.
   Nucl.\ Phys.\ \rf{44}{145}{86}].  
\defref{35} G.P. Korchemskii, ``Double Logarithmic Asymptotics in QCD'',
   Dubna preprint E2-88-600, and ``Sudakov Form Factor in QCD'', Dubna
   preprint E2-88-628.
\defref{36} P.V.  Landshoff, \pl\rf{169B}{69}{86}, and references therein.
   % axial gauge problems.  Need other papers? NO
\defref{37} P.J.  Doust and J.C.  Taylor, \pl\rf{197B}{232}{87}.  
   % Coulomb gauge problems.  Others? NO
\defref{38}
    D. Amati, R. Petronzio, and  G. Veneziano, \np\rf{B140}{54}{78} and
    \rf{B146}{29}{78};
    R.K. Ellis, H. Georgi, M. Machacek, H.D. Politzer, and G.G. Ross,
    \np\rf{B152}{285}{79};
    A.V. Efremov and A.V. Radyushkin, Teor.\ Mat.\ Fiz.\ \rf{44}{17}{80}
    [Eng.\ transl.: Theor.\ Math.\ Phys.\ \rf{44}{573}{81}], 
    Teor.\ Mat.\ Fiz.\ \rf{44}{157}{80} 
    [Eng.\ transl.: Theor.\ Math.\ Phys.\ \rf{44}{664}{81}],
    Teor.\ Mat.\ Fiz.\ \rf{44}{327}{80}
    [Eng.\ transl.: Theor.\ Math.\ Phys.\ \rf{44}{774}{81}]; % gauge theories
    Libby and Sterman, Ref.~\ref{11}.

\end{sloppypar}

\end{enumerate}

\end{document}

%% file: mclmath.tex
\input jccsym
\relax

%% file: jccsym.tex
\expandafter \ifx \csname JCCSYMTEX\endcsname \relax
   \def \TMPTMPTMP{}
\else
   \def \TMPTMPTMP{ }
\fi
\TMPTMPTMP

\def \fr#1#2{\ifinner \innerfr{#1}{#2}  \else \outerfr{#1}{#2}\fi}
\def \innerfr#1#2{\ifmmode \ndmathfr{#1}{#2}
           \else \textfr{#1}{#2}\fi}
\def \outerfr#1#2{\ifmmode \dmathfr{#1}{#2}
           \else \textfr{#1}{#2}\fi}
\def \ndmathfr #1#2{ \raise 0.4 ex\hbox{${\scriptstyle {#1 \over #2}} $} }
\def \dmathfr  #1#2{ {#1 \over #2} }
\def \textfr   #1#2{\raise 0.4 ex\hbox{${\scriptstyle {#1 \over #2}}$}}
\def\fun#1#2{\lower3.6pt\vbox{\baselineskip0pt\lineskip.9pt
  \ialign{$\mathsurround=0pt#1\hfil##\hfil$\crcr#2\crcr\sim\crcr}}}
\def\centeron#1#2{{\setbox0=\hbox{#1}\setbox1=\hbox{#2}\ifdim
\wd1>\wd0\kern.5\wd1\kern-.5\wd0\fi
\copy0\kern-.5\wd0\kern-.5\wd1\copy1\ifdim\wd0>\wd1
\kern.5\wd0\kern-.5\wd1\fi}}
\def\centerover#1#2{\centeron{#1}{\setbox0=\hbox{#1}\setbox
1=\hbox{#2}\raise\ht0\hbox{\raise\dp1\hbox{\copy1}}}}
\def\centerunder#1#2{\centeron{#1}{\setbox0=\hbox{#1}\setbox
1=\hbox{#2}\lower\dp0\hbox{\lower\ht1\hbox{\copy1}}}}
\def\etal{\hbox{\it et al.}}

\def \UNIT #1{{\ifmmode \UNITBODY {#1} \else $ \UNITBODY {#1} $\fi }}
\def \UNITBODY #1{\,{\rm #1}}

\def \al {\alpha_{\rm s}}
\def \alpi {{\alpha_s \over \pi}}

\def \MSbarbasic {\overline{{\rm MS}}}
\def \MSbar {\ifmmode \MSbarbasic \else $\MSbarbasic$\fi }

\def \st#1{\centeron{$#1$}{$/$}}

\def \bra#1|{\mathopen{\langle#1\,|}}  %Dirac bra arg. is placed in bra symbol
\def \braket#1|#2>{\langle#1\,|\,#2\rangle} %Dirac braket arg's in bra & ket
\def \ket#1>{\mathclose{|\,#1\rangle}}  %Dirac ket arg. is placed in ket symbol

\def \ltap{\;\centeron{\raise.35ex\hbox{$<$}}{\lower.65ex\hbox{$\sim$}}\;}
\def \gtap{\;\centeron{\raise.35ex\hbox{$>$}}{\lower.65ex\hbox{$\sim$}}\;}
\def \gsim{\mathrel{\gtap}}

\def \d#1{\mathop{{\rm d}#1}\nolimits}

\def \e{{\rm e}}
\def \epem{{\ifmmode \epembody \else $ \epembody $\fi }}
\def \epembody{ \e^+\e^- }
\def \half{\fr12}

\def \rf    #1#2#3{#1, #2 (19#3)}
\def \jour  #1{#1\ \relax}

\def \np    {\jour {Nucl.\ Phys.}}

\def \pl    {\jour {Phys.\ Lett.}}

\def \pr    {\jour {Phys.\ Rev.}}

\def \prl   {\jour {Phys.\ Rev.\ Lett.}}

\relax